\documentclass[usenatbib]{mnras}

\usepackage{graphicx}	
\usepackage{amsmath}	
\usepackage{amssymb}	
\usepackage{bm}		
\usepackage{soul}
\usepackage[flushleft]{threeparttable}
\usepackage{multirow}
\usepackage{Custom_Functions}

\defcitealias{kollmeier2014}{K14} 	
\defcitealias{danforth2016}{D16}	
\defcitealias{gaikwad2016}{Paper-I}	
\defcitealias{shull2015}{S15}		

\usepackage{xspace}
\newcommand{\viper}{{\xspace}{\sc viper}\xspace}	
\newcommand{\ltf}{{\xspace}{G2-LTF}\xspace}	
\newcommand{\htf}{{\xspace}{G2-HTF}\xspace}	
\newcommand{\gthree}{{\xspace}{\sc gadget-3}\xspace}	
\newcommand{\gtwo}{{\xspace}{\sc gadget-2}\xspace}	

\newcommand{\HI}{{\xspace}H{\sc i}\xspace}   		
\newcommand{\logNHI}{\,$\log{\rm N_{HI}}$\xspace}   		
\newcommand{\dlogNHI}{\,d$\log{\rm N_{HI}}$\xspace}   		
\newcommand{\GHI}{$\Gamma_{\rm HI}$\xspace}   				
\newcommand{\GTW}{$\Gamma_{\rm 12}$\xspace}   				

\newcommand{\lya}{\,Ly$\alpha$ }

\newcommand{\mpc}{\,Mpc}  		
\newcommand{\kmps}{\,km\,s$^{-1}$} 

\usepackage[usenames, dvipsnames]{color}
\definecolor{mycolor}{RGB}{0,128,0}
\definecolor{newaddcolor}{RGB}{255,0,255}

\usepackage[normalem]{ulem}

\title[Efficient adiabatic simulations of the high-$z$ IGM]{Efficient adiabatic hydrodynamical simulations of the high-redshift intergalactic medium}

\author[Gaikwad et.al]{Prakash Gaikwad$^{1}$\thanks{E-mail: \href{prakashg@ncra.tifr.res.in}{prakashg@ncra.tifr.res.in}}, 
Tirthankar Roy Choudhury$^{1}$,
Raghunathan Srianand$^{2}$,
\newauthor{and Vikram Khaire$^{1,3}$}
\\
$^{1}$National Centre for Radio Astrophysics, Tata Institute of Fundamental Research, Pune 411007, India \\
$^{2}$Inter-University Centre for Astronomy and Astrophysics (IUCAA), Post Bag 4, Pune 411007, India \\
$^{3}$Department of Physics, University of California, Santa Barbara, CA 93106, USA}

\date{}

\pubyear{2015}

\begin{document}
\label{firstpage}
\pagerange{\pageref{firstpage}--\pageref{lastpage}}
\maketitle


\begin{abstract}
We present a post-processing tool for \gtwo adiabatic simulations to model various observed properties of the \lya forest at $2.5 \leq z \leq 4$ that enables an efficient parameter estimation. In particular, we model the thermal and ionization histories that are not computed self-consistently by default in \gtwo. We capture the effect of pressure smoothing by running \gtwo at an elevated temperature floor and using an appropriate smoothing kernel. We validate our procedure by comparing different statistics derived  from our method with those derived using self-consistent simulations with \gthree. These statistics are:  line of sight density field power spectrum, flux probability distribution function, flux power spectrum, wavelet statistics, curvature statistics, \HI column density (${\rm N_{HI}}$) distribution function, linewidth ($b$) distribution and $b$ versus \logNHI scatter. For the temperature floor of $10^4$ K and typical signal-to-noise of 25, the results agree well within 20 percent of the self-consistent \gthree simulation. However, this difference is smaller than the expected $1\sigma$ sample variance for an absorption path length of $\sim 5.35$ at $z=3$.  Moreover for a given cosmology, we gain a factor of $\sim N$ in computing time for modelling the intergalactic medium under $N \gg 1$ different thermal histories. In addition, our method allows us to simulate the non-equilibrium evolution of thermal and ionization state of the gas and include heating due to non-standard sources like cosmic rays and high energy $\gamma$-rays from Blazars.  

\end{abstract}

\section{Introduction}
The \lya forest seen in the spectra of distant background QSOs trace the distribution of neutral hydrogen (\HI) in the universe at mildly non-linear overdensities  \citep[$\Delta \lesssim 10$,][]{miralda1996,bi1997,croft1997}.
Observed properties of the \lya forest are sensitive to fluctuations in the cosmic density and velocity fields and physical conditions like the temperature, turbulence and ionizing radiation prevailing in the intergalactic medium \citep[IGM;][]{cen1994,zhang1995,miralda1996,hernquist1996}. As a result, \lya forest has been used in the literature to constrain cosmological parameters such as $\Omega_{\rm m}$, $\Omega_{\rm b}$, $\sigma_{\rm 8}$, $n_{\rm s}$ \citep[see, e.g.,][]{viel2004a,viel2004b,mcdonald2005},
the neutrino mass \citep{palanque2015a,palanque2015b,yeche2017}, mass of warm dark matter particles \citep{narayanan2000,viel2005,viel2013b} and astrophysical parameters such as the IGM temperature $T_0$ at cosmic mean density and slope $\gamma$ of the temperature ($T$) - density ($\Delta$) relation \citep[$T = T_0 \Delta^{\gamma-1}$, hereafter TDR; ][]{schaye1999,schaye2000,zaldarriaga2001,mcdonald2001,theuns2000b,lidz2010,becker2011,
boera2014} and H~{\sc i} photo-ionization rate \citep[\GHI,][]{rauch1997,cooke1997,meiksin2004,becker2013,kollmeier2014,shull2015,gaikwad2017a,gaikwad2017b,viel2016,gurvich2017}. 

Usually constraining these parameters involves comparing different properties of the \lya forest derived from observed spectra with those from the simulated ones. Early simulations of \lya forest based on lognormal \citep{bi1992,bi1997,gnedin1996,trc2001} or the Zel’dovich approximation \citep{doroshkevich1977,mcgill1990}, although fast and capture the basic picture, failed to reproduce the quasi- and non-linear density fields accurately \citep{2002MNRAS.336..685V}
or washed out the small scale structures in the \lya forest \citep{gnedin1998}. 

By using cosmological $N$-body simulations \citep{hernquist1989,springel2005,oshea2005}, the \lya forest has been modelled in the past using (i) dark matter only simulations where baryons are assumed to follow the dark matter, and the temperature is assigned to the baryons assuming a power-law TDR \citep{mucket1996}, (ii) smoothed particle hydrodynamic (SPH) codes \citep{cen1994,zhang1995,hernquist1996,theuns1998a,dave1999,viel2004a} like \gtwo and \gthree\footnote{\gthree is not publicly available. However, Volker Springel has provided this code through private communication for our studies. \gthree has been frequently used for \lya forest studies  \citep[see e.g., ][]{becker2011,viel2016}} \citep{springel2001,springel2005,bolton2006}, (iii) grid based adaptive mesh refinement (AMR) code {\sc enzo} \citep{smith2011,shull2012b,enzo2014} and (iv) hybrid methods such as \lya Mass Association Scheme (LyMAS) in which moderate resolution dark matter only simulation is used after the calibration using high resolution but small volume hydrodynamic simulations \citep{peirani2014,sorini2016}. The main drawback of the dark matter only simulations is that it does not account for the smoothing of the baryonic density field due to finite pressure of the baryons, while these effects are self-consistently accounted for in the SPH and AMR based simulations. Interestingly, the \lya forest flux statistics from SPH and AMR simulations are shown to agree with each other to within 10 per cent accuracy \citep{regan2007}. These simulations can, in principle, incorporate different complex astrophysical processes such as the radiative heating, cooling, shocks, starbursts and AGN induced feedback processes \citep{kollmier2006,mcdonald2006,dave2010,schaye2010,viel2013}. 

While the current state of the art hydrodynamical simulations are extremely useful for probing the physical properties of the IGM, the computational expenses severely limit their usage for constraining the unknown model parameters and their associated errors. Various approaches have been introduced to keep the computational expense within manageable limits while exploring the large parameter space. For example, \citet{viel2006a,viel2009} begin by choosing a ``best-guess'' model and expand the statistical quantities under consideration (e.g., the flux power spectrum) in a Taylor series around this model. Their method requires calculating a limited number of derivatives which can be achieved by running only a few simulations around the best-guess model. The method of \citet{mcdonald2005} involves running simulations on a carefully chosen grid in the parameter space and then interpolating between these runs. Other methods include deriving scaling relations between different parameters from a limited number of hydrodynamical simulations which are useful for studying parameter degeneracies \citep{bolton2005,bolton2007,faucher2008c}. Since many of the parameters, particularly those related to the thermal state of the IGM, are poorly understood, obtaining robust constraints would require exploring a sufficiently wide range of parameter values. It is thus useful to develop newer methods of simulating the high-$z$ IGM that are efficient, flexible and at the same time sufficiently accurate. This forms one of the main motivation of this work.

In \citet{gaikwad2017a}, we have developed a ``Code for Ionization and Temperature Evolution'' {\sc (cite)} to estimate the temperature of the  SPH particles in the post-processing step of \gtwo by taking care of radiative cooling and heating effects. {\sc cite} allowed us to place good constraints on \GHI while efficiently exploring different thermal histories at low-$z$ ($z \leq 0.5$). While {\sc cite} works well for the low resolution simulation (gas particle mass $\delta m = 1.26 \times 10^7 \: h^{-1} \: {\rm M}_{\sun}$ and pixel size $\delta x = 48.8h^{-1}$ ckpc) as shown in \citet{gaikwad2017a}, the dynamical evolution of SPH particles at finite pressure is an important effect when we consider high resolution simulations (e.g. gas particle mass $\delta_m = 1.01 \times 10^5 \: {\rm M}_{\sun}$ and pixel size $\delta x = 9.77h^{-1}$ ckpc). In this article, we present a method to account for this effect by smoothing (in 3 dimensions) the density and velocity fields over a local Jeans scale.
We explore the consistency of our method with that from \gthree \citep[in which the thermal effects on the hydrodynamical evolution of baryonic particles are taken care of in a self-consistent manner]{springel2005} by comparing different \lya flux statistics frequently used in the literature. Our method (though approximate) is computationally less expensive and accurate enough to constrain physical parameters through a detailed exploration of possible parameter space. Our code is also flexible enough to incorporate effects such as non-equilibrium evolution of the ionization state of the gas and heating by non-standard sources like Blazars or cosmic rays etc.

This paper is organized as follows. In \S2, we describe the \gtwo and \gthree simulations used in this study. We discuss the method of simulating \lya forest in \S3. We show the consistency of our method with \gthree by comparing 8 different statistics in \S4. We summarize our results in \S5.  We use flat $\Lambda$CDM cosmology with parameters $(\Omega_{\Lambda}, \Omega_{\rm m}, \Omega_{\rm b}, h, n_s,\sigma_8,Y) \equiv (0.69, 0.31, 0.0486, 0.674, 0.96, 0.83, 0.24)$ consistent with \citet{planck2016}. The \HI photoionization rate (\GHI) expressed in units of $10^{-12} \: {\rm s}^{-1}$  is denoted as \GTW. Unless mentioned all the distances are expressed in comoving co-ordinates.

\section{Simulation}
\label{sec:simulation}

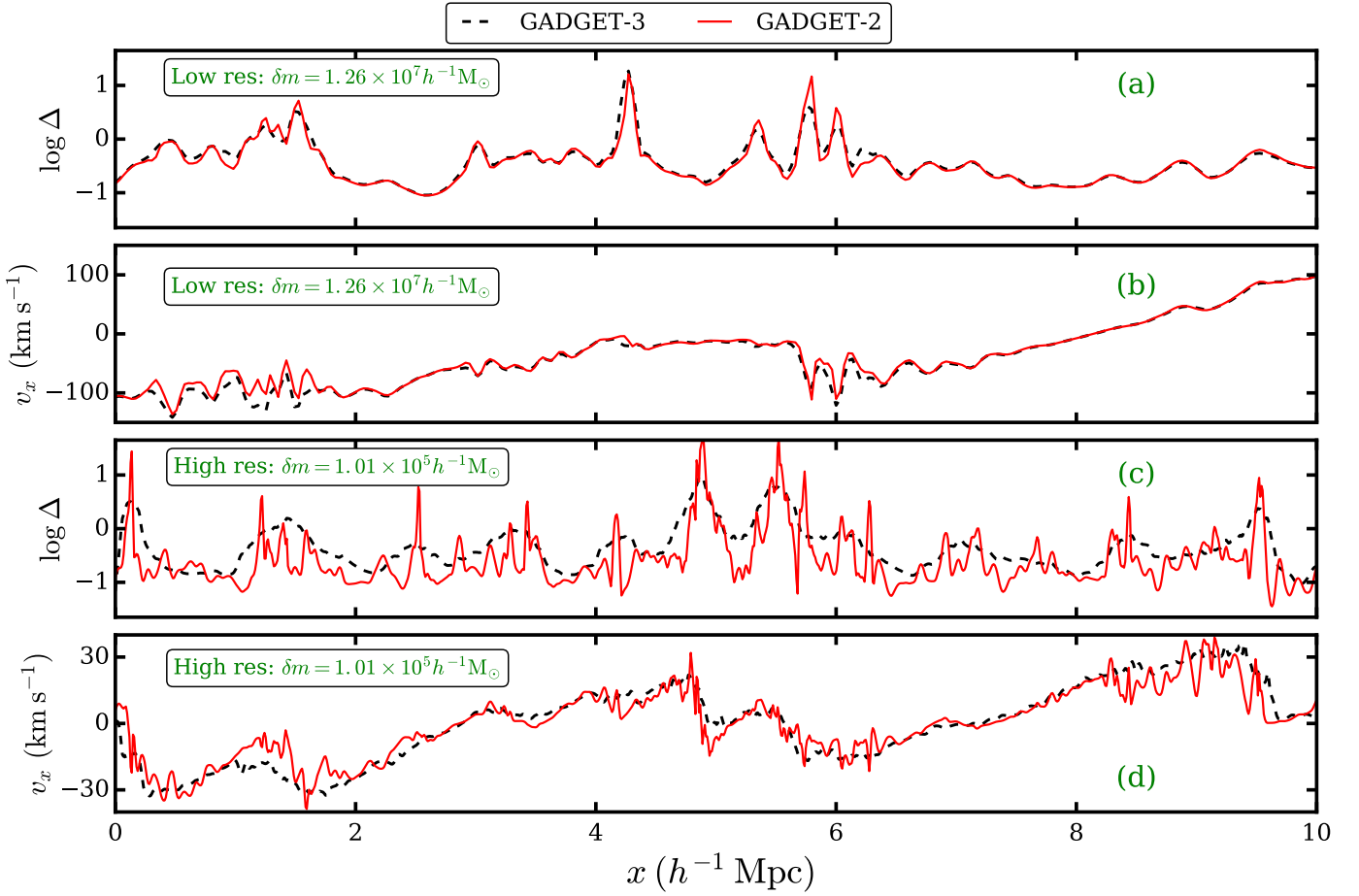
\InputFigCombine{box_size_effect.pdf}{180}{Panels (a) and (b) compare the line of sight density and velocity fields respectively from \gthree (black dashed curve) and \gtwo (red solid curve) simulations for a low resolution simulation box at $z=2.5$ (box size $L=50h^{-1}$ cMpc, gas particle mass $\delta m = 1.26 \times 10^7 \: h^{-1} \: {\rm M}_{\sun}$ and pixel size $\delta x = 48.8h^{-1}$ ckpc).  Panels (c) and (d) are same as panels (a) and (b) respectively except that these are obtained from high resolution simulation boxes at $z=2.5$ (box size $L=10h^{-1}$ cMpc, gas particle mass  $\delta_m = 1.01 \times 10^5 \: {\rm M}_{\sun}$ and pixel size $\delta x = 9.77h^{-1}$ ckpc) used in this paper. \gtwo models for low and high resolution boxes are performed with the temperature floor of $\sim 100$ K.}{\label{fig:box-size-effect}}

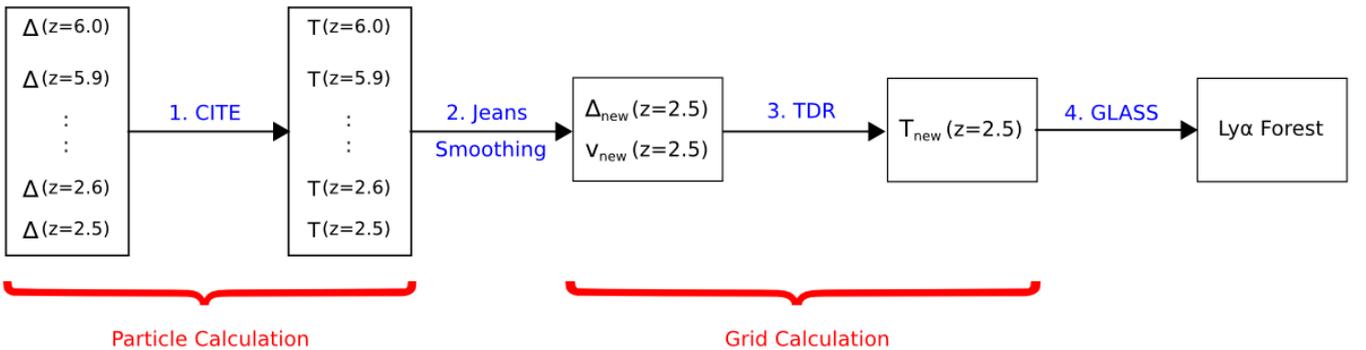
\InputFigCombine{Capture.png}{180}{Schematic diagram showing main steps adopted in our post-processing method of obtaining \lya forest spectra from {\sc gadget-2} taking into account radiative cooling and heating effects externally. The basic steps involved in our method are: (1) We calculate the temperature of each particle at each redshift using {\sc cite} and obtain the thermal history parameters $T_0$ and $\gamma$. (2) Given $T$ and $\Delta$ of particles, we apply pressure smoothing to get new $\Delta_{\rm new}$ and $v_{\rm new}$ on grids for a simulation box at a redshift of interest. (3) For this new $\Delta$ on grid points, we apply power-law TDR using thermal history parameters $T_0$ and $\gamma$ obtained in the previous step. (4) We calculate \lya optical depth from the simulation box using our routine {\sc glass}.}{\label{fig:flag-method}}

We use the publicly available \gtwo\footnote{http://wwwmpa.mpa-garching.mpg.de/gadget/} \citep{springel2005} to perform smoothed particle hydrodynamical simulations used in this study. The initial conditions are generated at $z=99$ using the publicly available {\sc 2lpt}\footnote{http://cosmo.nyu.edu/roman/2LPT/} code \citep{2lpt2012}. We use $1/30^{\rm th}$ of the mean inter-particle distance as the gravitational softening length. The \gtwo simulation does not include radiative heating and cooling of the SPH particles internally. As a result, the unshocked gas particles (in the low density regions) are evolved at very low temperature (the default value is $100$ K in \gtwo) and pressure. However, the simulation allows one to set the minimum allowed gas temperature (referred as temperature floor) to higher values. In this work, we perform two simulations of \gtwo: (i) \ltf with low temperature floor of $T=100$ K and (ii) \htf  with high temperature floor of $T = 10000$ K (corresponding to typical IGM temperatures due to photoheating). An unique identification number is assigned to each particle in \gtwo and is used for tracing its density and temperature evolution.
 
We also perform a \gthree simulation \citep[a modified version of the publicly available \gtwo code, see][]{springel2005}  with the same initial conditions as the \gtwo simulations discussed above. Unlike \gtwo, the \gthree simulation includes radiative heating and cooling of SPH particles internally for any given metagalactic UV background (UVB). We use \citet[][hereafter HM12]{haardt2012} UVB assuming ionization equilibrium in \gthree. To speed up the calculations, we run the simulations with QUICK\_LYALPHA flag  that converts particles with $\Delta > 1000$ and $T<10^5$ K into stars \citep{viel2004a} and removes them from subsequent calculations. None of our simulations (i.e., \gtwo or \gthree) include AGN feedback, stellar feedback or outflows in the form of galactic wind. The details of our simulations are listed in Table \ref{tab:sim-details}.

\begin{table*}
\caption{Details of our simulations described in \S\ref{sec:simulation}}
\begin{threeparttable}
\centering
\begin{tabular}{lccc}
\hline \hline
Model & \gthree & \ltf & \htf \\
\hline \hline
N-body code & \gthree &{\sc gadget-2} & {\sc gadget-2}\\
Initial redshift\tnote{1} & 99 & 99 & 99\\
Box size ($h^{-1}$ c\mpc)& 10 & 10 & 10  \\
Number of particles & $2 \times 512^3$ & $2 \times 512^3$ & $2 \times 512^3$ \\
UVB\tnote{2} & HM12 & HM12 & HM12\\
Ionization evolution\tnote{2} & Equilibrium & Equilibrium & Equilibrium\\
$T$ and $\Delta$ evolution & Internal & Post-process (\sc cite) & Post-process (\sc cite)\\
SFR Criteria\tnote{3} & QUICK\_LYALPHA & $-$ & $-$\\
Output redshifts & $6.0,5.9,\cdots,2.0$ & $6.0,5.9,\cdots,2.0$ & $6.0,5.9,\cdots,2.0$\\
Temperature floor\tnote{4} & $-$ & $100$ K & $10000$ K\\
Smoothing kernel type\tnote{5} & SPH & Modified & Modified\\
& $W(r,h)$ & $W^{\prime}(r,h,L_j)$ & $W^{\prime}(r,h,0.66 \times L_J)$\\
Gas particle mass ($\delta m$)\tnote{6} & $1.01 \times 10^5 \: h^{-1} \: {\rm M}_{\sun}$  & $1.01 \times 10^5 \: h^{-1}\: {\rm M}_{\sun}$ & $1.01 \times 10^5 \: h^{-1} \: {\rm M}_{\sun}$ \\
Pixel size ($\delta x$)\tnote{7} & $9.77 h^{-1}$ ckpc  & $9.77 h^{-1}$ ckpc & $9.77 h^{-1}$ ckpc \\
\hline \hline
\end{tabular}
\begin{tablenotes}
\item[1] All simulations (i.e. \gthree, \ltf and \htf) are performed using same initial condition. 
\item[2] The default run of \gthree solves equilibrium ionization evolution equation using HM12 UVB.
\item[3] The QUICK\_LYALPHA flag in \gthree converts gas particles with $\Delta > 1000$ and $T < 10^5$K in to stars.
\item[4] The minimum allowed temperature of the gas particle in simulation is set by the temperature floor.
\item[5] To account for pressure smoothing in \ltf and \htf model, the smoothing kernel is modified by convolving SPH kernel with Gaussian kernel of pressure smoothing in the post-processing step. The pressure smoothing is self-consistently accounted for in the default run of \gthree model.
\item[6] The gas particle mass refers to the minimum mass of baryon particles in our model runs.
\item[7] The pixel size refers to the scale on which quantities  (like $\Delta$, $v$ and $T$) are gridded when computing the spectra.
\end{tablenotes}
\end{threeparttable}
\label{tab:sim-details}
\end{table*}

\section{Method}
\label{sec:method}

The \lya optical depth is calculated by evaluating the overdensity ($\Delta$), temperature ($T$) and velocity ($v$) on grid points along a given sightline in the simulation box. Unlike \gthree, the TDR obtained in \gtwo is not realistic as the radiative heating and cooling terms are not incorporated. At moderate to low resolution, the overdensity and velocity fields from \gtwo matches well with those from \gthree  as shown in panel (a) and (b) of Fig. \ref{fig:box-size-effect}. This resolution (gas particle mass $\delta m = 1.26 \times 10^7 \: h^{-1} \: {\rm M}_{\sun}$, pixel size $\delta x \sim 48.8h^{-1}$ ckpc) is appropriate for low-$z$ ($z<0.5$) \lya forest studies with instruments like the HST-COS \citep{gaikwad2017a,gaikwad2017b}. However \gtwo does not capture the effect of finite gas pressure in the hydrodynamical evolution of the photoionized gas. This effect becomes important at smaller scales probed well in high resolution spectra (gas particle mass $\delta m = 1.01 \times 10^5 \: h^{-1} \: {\rm M}_{\sun}$, pixel size $\delta x \sim 9.77h^{-1}$ ckpc) typically used in the \lya forest studies at high-$z$ ($z>1.6$). This is illustrated in the panel (c) and (d) of Fig. \ref{fig:box-size-effect} where the density and velocity fields obtained in \gthree can be seen to be smooth as compared to those in \gtwo. Our method of evolving the gas temperature using \gtwo + {\sc cite}, as discussed in \citet{gaikwad2017a}, does not account for the effect of finite gas pressure on the evolution of density and velocity fields. 

In this work, we present a method to account for the effect of gas pressure in \gtwo + {\sc cite} for high resolution \lya forest simulations. Fig. \ref{fig:flag-method} shows the outline of our procedure whose main steps are as follows: (1) First we estimate the temperature of the \gtwo particles accounting for the radiative heating and/or cooling \citep{gaikwad2017a}. Depending on the requirements of the problem, the ionized fraction can be calculated either under ionization equilibrium or non-equilibrium conditions. (2) We then calculate the Jeans length for each particle assuming the particles to be in local hydrostatic equilibrium \citep{schaye2001}. We smooth the density field by modifying the SPH kernel suitably to account for pressure smoothing. (3) We then use the TDR to calculate the temperature on the grids \citep{hui1997} for particles that do not go through any shock heating. (4) Finally we calculate the \lya optical depth using the density, velocity and temperature along the sightline \citep{trc2001}. We discuss all these steps in more details below. 

\subsection{Temperature evolution in \gtwo using {\sc cite}:}

\InputFigCombine{EoS_Particle.pdf}{170}{TDR of the SPH particles from \gthree (left panel), \ltf (middle panel) and \htf (right panel) at $z=2.5$. The temperature in the \ltf and \htf models are obtained in the post-processing step of \gtwo using {\sc cite} (see \S \ref{sec:method}). The magenta dashed vertical lines show bins in $\log \Delta$. We calculate median $T$ (black stars) in each of these $\Delta$ bins and fit a power-law, $T=T_0 \: \Delta^{\gamma-1}$, to obtain $T_0$ and $\gamma$. The resulting TDR is shown by black dashed line. In the case of \gthree we use {\sc quick\_lyalpha} flag under which gas particles with $T<10^5$ K and $\Delta > 1000$ are converted into stars and got removed from subsequent calculations. No such star formation criteria is applied in \ltf and \htf models (see Appendix \ref{app:star-formation} for more details). The colour scheme represents density of points in logarithmic unit.}{\label{fig:eos-comparison}}

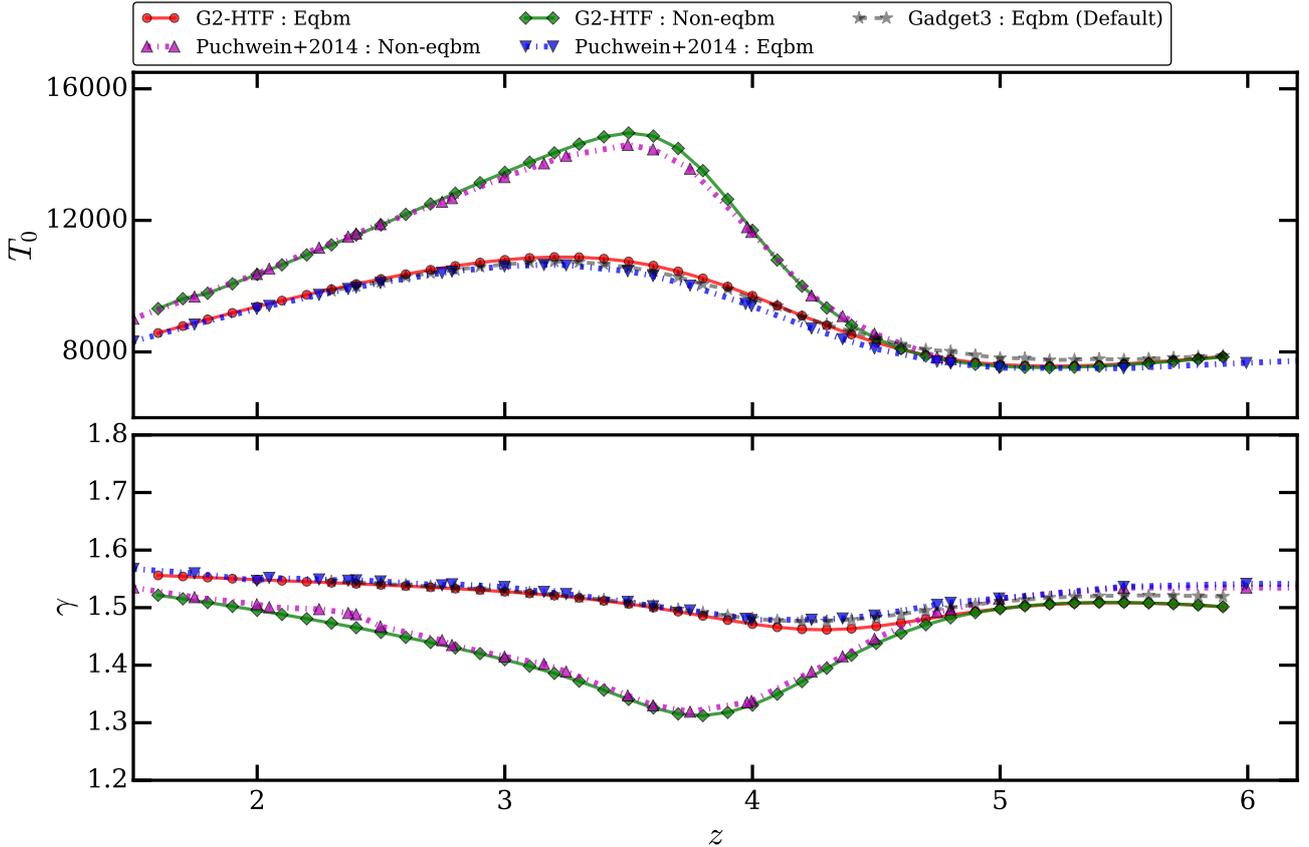
\InputFigCombine{Thermal_history.pdf}{170}{Comparison of redshift evolution of the thermal history parameters ($T_0$ and $\gamma$) from our \htf with \gthree (gray stars) simulations and that of \citet[][magenta up-triangles for non-equilibrium and blue down-triangles for equilibrium ionization evolution]{puchwein2015}. {\sc cite} is started at $z=6.0$ with initial conditions $T_0 = 7920$ K and $\gamma=1.52$ same as those obtained in \gthree at that redshift (see \S\ref{sec:method} for details). For \htf simulations we run {\sc cite} using equilibrium (red filled circles) and non-equilibrium (green diamonds) ionization condition influenced by the same UVB. Note that the default version of \gthree solves equilibrium ionization evolution equation.}{\label{fig:T0-gamma-evolution}}

We  evolve the temperature of the particles in \gtwo using {\sc cite} \citep[as discussed in details in][]{gaikwad2017a}. For completeness, here we briefly discuss the steps involved. We solve the temperature evolution equation for each particle in the post-processing step of \gtwo using
\begin{equation}\label{eq:temperature-evolution}
\frac{dT}{dt} = -2HT + \frac{2T}{3\Delta} \: \frac{d \Delta}{dt} +  \frac{dT_{\rm shock}}{dt}  +   \frac{dT_{\rm IE}}{dt} + \frac{dT_{\rm other}}{dt} \;\; .
\end{equation}
The five terms on the right hand side of above equation represents, respectively, rate of cooling due to Hubble expansion, adiabatic heating or cooling arising from change in density of particles, change in temperature due to shock heating, change in temperature due to change in internal energy per particle and change in temperature due to other heating/cooling processes (such as photo-heating, cosmic ray heating, radiative cooling). We use {\sc cite} to calculate the last two terms on right hand side of Eq. \ref{eq:temperature-evolution} as they are not self-consistently computed in \gtwo. The actual implementation is as follows.

\begin{enumerate}

\item At the initial redshift (taken to be $z_1=6.0$ in this work), we assume a given power-law TDR. In this paper, we choose $T_0 = 7920$ K and $\gamma=1.52$ in order to match those obtained in \gthree at the same redshift for HM12 UVB. We then compute the actual temperature of a gas particle using following prescription: If a particle is shock heated in recent times (i.e., within a time scale corresponding to $\delta z=0.1$), then the temperature of the particle will not be updated by {\sc cite}. Otherwise we assume the particle temperature to be following the above mentioned power-law TDR. At the initial redshift, we solve equilibrium ionization evolution equation assuming HM12 UVB to calculate various ion fractions of H and He. 

\item Given the ion fractions and the temperatures, it is straightforward to calculate last two terms on the right hand side of Eq. \ref{eq:temperature-evolution} for subsequent time steps. For this, we use the photo-heating rates of HM12 UVB model.

\item To obtain the temperature of the particles in the next time step ($z_2 = z_1 - \Delta z$)\footnote{In all simulations, we have stored the \gtwo snapshots between $z=6$ to $2$ with a redshift interval of 0.1 (see \S\ref{sec:simulation}). In {\sc cite}, we divide the time-step between two neighbouring redshifts into 100 smaller steps for numerical stability (i.e., $\Delta z = 0.001$) and interpolate all the relevant quantities in the intermediate time-steps.}, we first check if the particle is shock heated in recent times (i.e., within a time scale corresponding to $\delta z=0.1$). If the particle is not shock heated, then we neglect the third term on the right hand side of Eq. \ref{eq:temperature-evolution}. Otherwise we solve the same Eq. \ref{eq:temperature-evolution} accounting all the five terms.
  
\item For redshift $z_2$, we solve equilibrium (or non-equilibrium, if desired) ionization evolution equations to calculate various ion fractions.

\item We repeat the steps (ii)-(iv) to obtain the temperature of the particle at subsequent  redshifts.

\end{enumerate}

Fig. \ref{fig:eos-comparison} shows comparison of TDR  of SPH particles obtained from \gthree (left panel), \ltf (middle panel) and \htf (right panel) simulations at $z=2.5$. Qualitatively, the TDR from \ltf and \htf (after processing through {\sc cite}) is remarkably similar to that from \gthree. The differences at $\Delta > 1000$ and $T<10^5$ K can be attributed to the QUICK{\_}LYALPHA flag employed in \gthree  (see Appendix \ref{app:star-formation} for more details). For each model, we calculate median temperature (black star points) in $\log \Delta$ bins with centres at $-0.375,-0.125,0.125,0.375$ and bin width $0.125$ (indicated by magenta dashed vertical lines). We then fit power law relation $T=T_0 \: \Delta^{\gamma-1}$ to obtain the best fit $T_0$ and $\gamma$ \citep{hui1997,mcdonald2005}. The fitted TDR is shown by black dashed line in each panel. The values of $T_0$ and $\gamma$ are also indicated in each panel. It is clear that they are similar within $2.5$ percent. 

Fig. \ref{fig:T0-gamma-evolution} shows the redshift evolution of best fit $T_0$ (top panel) and $\gamma$ (bottom panel) for \htf , \gthree and \citet{puchwein2015} models for equilibrium and non-equilibrium ionization evolution cases. The evolution of $T_0$ and $\gamma$ obtained from {\sc cite} for the equilibrium ionization case is remarkably similar to those obtained from the \gthree run and \citet{puchwein2015}\footnote{The differences between the values of $T_0$ and $\gamma$ calculated from \ltf and \htf are less than 0.1 per cent.}.
As mentioned earlier, we can also solve for non-equilibrium ionization evolution equation using {\sc cite}. The $T_0$ and $\gamma$ evolution for non-equilibrium case from \citet[][magenta dashed curve]{puchwein2015} is also consistent with those from \htf with the maximum difference being less than 2.5 per cent (at $z \sim 3.5$). Since the default version of \gthree solves the ionization evolution equation under equilibrium conditions, hereafter we restrict our discussions to the models with equilibrium ionization as we will use \gthree as our reference. While {\sc cite} reproduces the $T_0$ and $\gamma$ evolution well, the issues related to small scale density and velocity field (demonstrated in Fig. \ref{fig:box-size-effect}) still need to be addressed.

\subsection{Jeans length of SPH particle in \gtwo :}

In this section, we explore the possibility of using local pressure smoothing in the \gtwo simulations to reduce the shortcomings highlighted in panels (c) and (d) of Fig. \ref{fig:box-size-effect}. We  choose to smooth the density field in \ltf or \htf on the scales of Jeans length of the particles to account for the pressure smoothing. Assuming the \lya absorbers to be in local hydrostatic equilibrium, \citet{schaye2001} has shown that the Jeans length can be obtained by equating dynamical time with sound  crossing time and is given by,
\begin{equation}{\label{eq:jeans-length}}
\frac{L}{1 \: {\rm kpc}} \sim 0.52  \:  \bigg[\frac{T}{10^4 \: {\rm K}} \:  \frac{1-Y}{0.76} \: \frac{f_g}{0.16} \: \frac{1 \: {\rm cm}^{-3}}{n_H} \: \frac{0.59}{\mu} \bigg]^{1/2}
\end{equation}
where, $T$ is temperature, $n_H$ is number density of H, $Y$ is He fraction by mass, $\mu = 4 / (8-5Y)$ is the mean molecular weight and $f_g$ is fraction of total mass in gas phase. For the scales of interest here $f_g$ is close to its universal value $\Omega_{\rm b} / \Omega_{\rm m} \sim 0.16$. It should be emphasized that the Jeans length depends on the density and temperature and hence is different for different particles. For the same reason, it is different for the same particle at different epochs. The above equation is not valid for \lya absorbers with characteristic densities smaller than the cosmic mean \citep[$\Delta \sim 1$,][]{schaye2001}. 
Hence we ignore the pressure smoothing for such particles and retain only the SPH smoothing. We now explain how the effect of pressure smoothing is incorporated in \ltf or \htf by modifying the SPH kernel.

\subsection*{Smoothing kernel :}

The estimate of a quantity $f$ at any grid point $i$ in the SPH formulation \citep{monaghan1992,springel2005} is given by,  
\begin{equation}{\label{eq:grid-estimate}}
f_i = \sum\limits_{j} \; f_j \; \frac{m_j}{\rho_j} \; S_{ij}
\end{equation}
where the summation is performed over all particles. The quantities $m_j$, $\rho_j$, $f_j$ are the mass, density and value of the quantity $f$ of $j^{th}$ particle, respectively. The quantity $f$ could be overdensity ($\Delta$), temperature ($T$) or any component of the velocity ($v$). The smoothing kernel, $S_{ij}$,  has units of inverse of volume and in general depends on the distance ($r_{ij}$) between $i^{th}$ grid point and $j^{th}$ particle. It is necessary for $S_{ij}$ to satisfy the following normalization condition in order to conserve the quantity $f$ (in particular mass) in SPH formulation \citep{monaghan1992},
\begin{equation}{\label{eq:kern-norm}}
\int\limits_{\mathcal{V}} \; S_{ij} \; d \bm{r} = 1
\end{equation}
where the integration is over volume $\mathcal{V}$. \\

We use the following smoothing kernels for various simulations,
\begin{equation}{\label{eq:smth-kenrel-cases}}
    S_{ij} \equiv
\begin{cases}
    \;\; W(r_{ij},h_j),&  {\rm For \; \textsc{gadget-3}}\\ 
    \;\; W^{\prime}(r_{ij},h_j,1 \times L_j),& {\rm For \; \textsc{\ltf}} \\
    \;\; W^{\prime}(r_{ij},h_j,0.66 \times L_j),& {\rm For \; \textsc{\htf}} \\
\end{cases}
\end{equation}
where $h_{j}$ and $L_j$ are smoothing length and Jeans length (given by Eq. \ref{eq:jeans-length}) of the $j^{th}$ particle respectively.

The smoothing kernel used for \gthree is same as SPH kernel given in \citet{springel2005} and has following form, 
\begin{equation} \label{eq:sph-kernal}
    W(r,h) = W_0
\begin{cases}
    \;\; 1 - 6 \bigg( \frac{r}{h} \bigg)^2 + 6\bigg( \frac{r}{h} \bigg)^3,&  0 \leq \frac{r}{h} \leq \frac{1}{2}\\
    \;\; 2\bigg( 1 - \frac{r}{h} \bigg)^3, & \frac{1}{2} \leq \frac{r}{h} \leq 1 \\ 
    \;\; 0, & \frac{r}{h} > 1
\end{cases}
\end{equation}
where $W_0 = 8 / (\pi h^3)$ is normalization constant of SPH kernel. 

The pressure smoothing can be well approximated by a Gaussian \citep{gnedin1998,girish2015}. Hence we modify the smoothing kernel by convolving SPH kernel with Gaussian kernel of pressure smoothing
\begin{equation}{\label{eq:smth-kern-js}}
\begin{aligned}
W^{\prime}(r,h,\sigma) &=  \int \: {\rm d}^3\bm{x_1} \: W(r_1,h) \; G(|\bm{r}-\bm{x_1}|,\sigma) \;  \\
\end{aligned}
\end{equation}
where the Gaussian kernel is assumed to be isotropic and is given by
\begin{equation}{\label{eq:gauss-kern-js}}
\begin{aligned}
G(|\bm{r}-\bm{x_1}|,\sigma) &= \frac{1}{(2 \pi \sigma^2)^{3/2}} \;\; {\rm exp} \bigg[{-\frac{|\bm{r}-\bm{x_1}|^2}{2\sigma^2}} \bigg] \\
&= \frac{1}{(2 \pi \sigma^2)^{3/2}} \;\; {\rm exp} \bigg[{-\frac{(r^2 + r_1^2 - 2 \: r \: r_1 \: \mu)}{2 \: \sigma^2}} \bigg] \\
\end{aligned}
\end{equation}
with $\mu$ being the cosine of the angle between $\bm{r}$ and $\bm{x_1}$ and $\sigma$ the width of the Gaussian which in turn depends on the Jeans length. 
At this point let us highlight some of the key properties of $W^{\prime}(r,h,\sigma)$ which are relevant for our calculations:
\begin{itemize}
\item Both $W(r,h)$ in Eq. \ref{eq:sph-kernal} and $W^{\prime}(r,h,\sigma)$ in Eq. \ref{eq:smth-kern-js} satisfy the normalization condition given in Eq. \ref{eq:kern-norm}.
\item The kernel in Eq. \ref{eq:smth-kern-js} does not have a closed form analytic solution, hence we need to calculate it numerically (See Appendix \ref{app:sph-convolution-expression} for more details).
\item Unlike $W(r,h)$, $W^{\prime}(r,h,\sigma)$ does not have a compact support as the Gaussian is non-zero at large distances. Hence we put a cut-off such that if distance between particle and grid is more than $h + 3 \sigma$, the contribution of $W^{\prime}(r,h,\sigma)$ is zero. Mathematically,
\begin{equation} \label{eq:smth-kernal}
    W^{\prime}(r,h,\sigma) = 
\begin{cases}
    \;\; W^{\prime}(r,h,\sigma) &  0 \leq r \leq (h + 3 \sigma)\\
    \;\; 0, & r > (h + 3 \sigma) \;\;\;\;\; .
\end{cases}
\end{equation}
We find that this cut-off does not have any significant effect on the density, velocity or temperature estimates as long as it is taken to be $\geq h + 3 \sigma$.
\item The amount of pressure smoothing in Eq. \ref{eq:smth-kern-js} is decided by the width $\sigma$ of the Gaussian. The SPH particles in \htf are evolved at relatively high temperature ($T \sim 10^4 $ K) and pressure as compared to \ltf ($T \sim 100$ K). It can be shown that the additional pressure smoothing length required in \htf model is factor $\sim 0.66$  times the smoothing length for the model \ltf (see Appendix \ref{app:jeans-length} for details).
\item This way of modifying smoothing kernel and estimating quantities along sightlines allow us to account for two important effects: (i) the variation in pressure smoothing for different particles at any epoch and (ii) the evolution of pressure smoothing scale for any particle at different epochs. Note that the pressure smoothing experienced by a particle in the \gthree simulation depends on the whole thermal history and not only on the present temperature as we do in our case \citep{lukic2015,girish2015}. However, as we will discuss later, running the \gtwo with high temperature floor captures (on an average) the pressure broadening arising from thermal history effects reasonably well.
\end{itemize}


\subsection{Estimation of the temperature field on a grid: }

After calculating the overdensity ($\Delta$) and velocity field ($v$) on grids along a given sightline using Eqs \ref{eq:grid-estimate}-\ref{eq:gauss-kern-js}, we can also estimate the temperature ($T$) along the same sightline using the same equations. However, the resultant TDR is not a power-law any more. This is because the temperature of the particle from {\sc cite} in the first step is calculated using \gtwo density field that does not incorporate the pressure smoothing. Hence we need to recalculate the temperature corresponding to the new density field with the pressure smoothing incorporated. In principle, we can again use {\sc cite} on the new smoothed density field and calculate the temperature. However, we find that this is computationally expensive because we need to calculate the smoothed density field on the grid along the sightline for all redshifts i.e. $z=6$ to $2$ with a $\Delta z =0.1$. Hence we adopt a simplified approach of applying power-law TDR \citep{hui1997,trc2001} 
\begin{equation} \label{eq:grid-eos}
    T = 
\begin{cases}
    \;\; T_0 \; \Delta^{\gamma-1}, &  \Delta \leq 10\\
    \;\; T_0 \; 10^{\gamma-1}, & \Delta > 10 \\
    \;\; T_{\rm shock}, & T_{\rm shock} > T
\end{cases}
\end{equation}
where $T_0$ and $\gamma$ are obtained from fitting the TDR for particles in our simulation box at the redshift of our interest as explained in Step (1) (also see Fig. \ref{fig:T0-gamma-evolution}). The last relation implies that if a particle is shock heated (or has temperature higher than that predicted by the TDR) then its temperature is not updated. We have confirmed that this approach produces consistent results with those obtained by running {\sc cite} on the new density field.

\subsection{\lya transmitted flux:} 

We have developed a module for ``Generating Ly-Alpha forest Spectra in Simulations'' {\sc (glass)} to calculate the \lya transmitted flux that has signal-to-noise ratio (SNR) and spectral resolution similar to the typical observational data used in the \lya forest studies.  The basic steps involved in {\sc glass} \citep{trc2001,hamsa2015,gaikwad2017a} are as follows: 
\begin{enumerate}
\item We determine the \HI number density ($n_{\rm HI}$) at any grid point from the baryonic density field ($\Delta$) assuming the gas to be optically thin and in photoionizing equilibrium with the UVB. The \HI photoionization rate (\GHI) is a free parameter. Throughout this paper we consider models with a fixed value \GHI $ = 10^{-12} \; {\rm s}^{-1}$ \citep{becker2013} for simplicity. 
\item  We calculate the \lya optical depth ($\tau$) along a line of sight from $n_{\rm HI}$ field by accounting for peculiar velocity, thermal and natural broadening effects.
\item The \lya transmitted flux is given by $F = {\rm e}^{-\tau}$.
\item When comparing with observations, the \lya flux field is linearly interpolated to match the wavelength sampling of observations.
\item The \lya flux field is then convolved with line spread function (LSF) of the spectrograph used in the observation. In this work we assume that the LSF is a Gaussian with a full width at half maximum, FWHM $\sim 7$ \kmps, typical of UVES or HIRES spectra.
\item Finally we add Gaussian random noise corresponding to a typical SNR=25 similar to what has been frequently achieved in echelle spectrographic observations with VLT and KECK that are used for \lya forest studies.
\end{enumerate}

\InputFigCombine{Slice.pdf}{170}{Slices from a simulation box having a width $\sim 10$ ckpc at $z=2.5$ for \gthree (top), \ltf (middle) and \htf (bottom). Left, middle and right panels in each row show overdensity ($\log \Delta$), velocity component ($v_x$) along $x$ axis and temperature ($\log T$) field respectively. The colour scheme represents density of points in logarithmic unit. We shoot a sightline parallel to $x$ axis through simulation box in each model as shown by horizontal dashed line in each panel. The extracted $\log \Delta$, $v_x$ and $\log T$ along these sightlines are plotted in Fig. \ref{fig:los-comparison} .}{\label{fig:slice-simulation}}

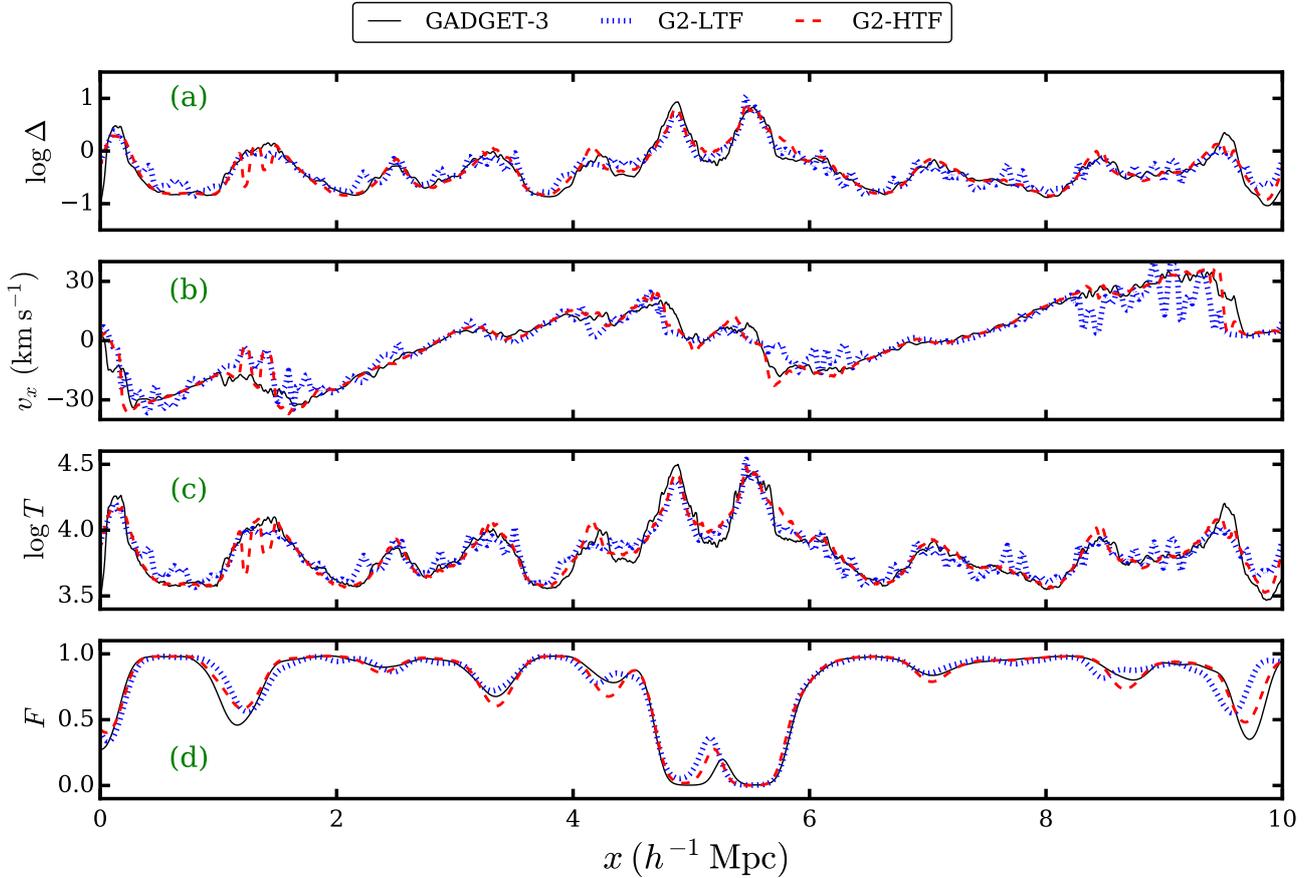
\InputFigCombine{LOS_comparison.pdf}{170}{Comparison of line of sight overdensity (panel (a), $\log \Delta$), velocity (panel (b), $v_x$ in \kmps), temperature (panel (c), $\log T$) and \lya transmitted flux (panel (d), $F$) for \gthree (black solid line), \ltf (blue dotted line) and \htf (red dashed line) from a simulation box at $z=2.5$ shown in Fig. \ref{fig:slice-simulation}. The \lya transmitted flux is not convolved with any LSF and no noise is added to the flux.}{\label{fig:los-comparison}}

A comparison of slices (having a width of $10$ ckpc) of the overdensity ($\log \Delta$), line of sight velocity (along $x$ axis, $v_x$)  and temperature ($\log T$) fields on grids from a simulation boxes at $z= 2.5$ are shown in Fig. \ref{fig:slice-simulation}. The top, middle and bottom rows show slices from \gthree, \ltf and \htf simulations respectively. The $\log \Delta$, $v_x$ and $\log T$ fields are sharper in the \ltf model (in particular in low density regions) as compared to those of \gthree model. On the other hand the $\log \Delta$, $v_x$ and $\log T$ fields from \htf model resembles close to those from \gthree. We shoot a sightline through each of these slices as shown by horizontal dashed line and extract the $\log \Delta$, $v_x$ and $\log T$ fields as shown in panel (a), (b) and (c) of Fig. \ref{fig:los-comparison} respectively. The line of sight $\log \Delta$, $v_x$ and $\log T$ fields from \ltf and \htf are very similar to those from \gthree. However, in general the variations in these fields for \ltf model are slightly more compared to those of \gthree and \htf models.  The panel (d) of Fig. \ref{fig:los-comparison} shows the \lya transmitted flux calculated along sightlines shown in Fig. \ref{fig:slice-simulation} for \gthree, \ltf and \htf models. Visually the \lya transmitted fluxes from different models are similar, despite subtle differences seen in $\log \Delta$, $v_x$, $\log T$ fields between these models. The \lya transmitted flux shown in this example is not convolved with LSF and is free of noise.

To perform a quantitative comparison of the \lya forest spectra extracted from different models, we identify eight statistics that are frequently used in the literature. We shoot random sightlines through the simulation and splice together the lines of sight in such a way that it covers a redshift path $z \pm 0.05$, where $z=2.5,3.0,3.5,4.0$ are redshifts of the simulation box\footnote{We do not splice together the lines of sight for FPS estimation.}. Each \lya forest spectrum has a path length of $\sim 50$ cMpc. Following \citet{rollinde,gaikwad2017a,gaikwad2017b}, we generate a mock sample of $N_{\rm spec} = 20$ \lya forest spectra for the \gthree, \ltf and \htf models. Each mock sample covers path length of $\sim 1000 h^{-1}$ cMpc (corresponding dimensionless absorption path length is $X \sim 5.35$ at $z=3$)\footnote{The dimensionless absorption path length is defined as $dX = dz\:  (1+z)^2 \:  \frac{H(0)}{H(z)}$ where $H(z)$ is hubble parameter at $z$ \citep{bahcall1969}.}.  This path length is similar to the path length covered in the \lya forest studies by \citet[][see their Table 3]{becker2011}. We repeat the procedure by choosing different random sightlines and generate $N=100$ such mock samples. The collection of $N$ mock samples constitute a ``mock suite'' that consists of $N \times N_{\rm spec} = 2000$ simulated spectra. Thus total path length covered in mock suite is $\sim 10^5$ cMpc. We estimate the covariance matrix for different statistics using the simulated spectra. 

\section{Results}
\label{sec:result}

We now compare different properties of the \lya forest generated from \ltf, \htf and \gthree simulations using eight statistics, namely, (i) the line of sight baryonic density field ($\delta = \Delta -1$) power spectrum (DPS), (ii) the flux probability distribution function (FPDF), (iii) the flux power spectrum (FPS), (iv) the wavelet statistics, (v) the curvature statistics, (vi) the column density distribution function (CDDF), (vii) the line width ($b$) distribution function and (viii) the $b$ vs \logNHI scatter plot. The statistics (i)-(v) are obtained assuming \lya transmitted flux to be a continuous field whereas, the statistics (vi)-(viii) are based on parameters derived using Voigt profile decomposition of \lya forest. For this purpose we use our automatic Voigt profile fitting code \viper described in full detail in \citet{gaikwad2017b}. 

\subsection{Line of sight density power spectrum (DPS)} 
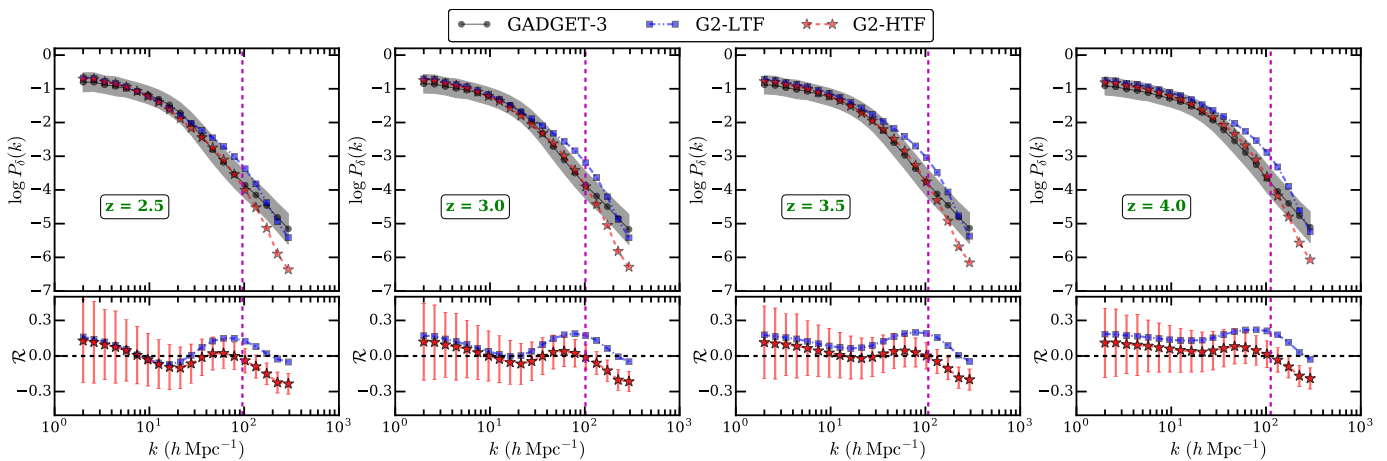
\InputFigCombine{Delta_PS.pdf}{180}{Top left panel shows the comparison of the line of sight density field power spectrum obtained from \gthree (black circle), \ltf (blue squares) and \htf (red stars) models at $z=2.5$. The gray shaded region represents the $1\sigma$ uncertainty (diagonal elements of covariance matrix given in Eq. \ref{eq:stat-cov}) on the DPS from \gthree. The Fourier modes with $k > k_{\rm cutoff}$ ($k_{\rm cutoff}$ varies with redshift e.g., $k_{\rm cutoff}(z=3)\sim 101h$ Mpc$^{-1}$, magenta dashed vertical line) are not be probed by current observations due to limited velocity resolution ($\sim 7$ \kmps) of the spectra. The residuals ($\mathcal{R}$, see Eq. \ref{eq:residual}) between the \ltf, \htf models with respect to the \gthree model at $z=2.5$ are shown in bottom left panel. The error-bars shown in this panel represent the $1\sigma$ uncertainties generated from a path length of $1000 h^{-1}$ cMpc (corresponding dimensionless absorption path length is $X \sim 5.35$ at $z=3$)  which correspond to the gray shaded region in the top left panel. The DPS from \ltf and \htf models are within 12 and 18 percent of \gthree model at scales $k < k_{\rm cutoff}$. The other panels show similar comparison at $4$ different redshifts that are identified in each panel.}{\label{fig:delta-ps-results}}

The density field power spectrum is not a directly measurable quantity but it influences all the observable quantities of \lya forest.
We calculate the power spectrum of the 1D density fluctuations along the line of sights using sightlines of comoving length equal to the simulation box size $10\:h^{-1}$ cMpc. 
This is done by computing the Fourier transform $\delta(k)$ of the density field $\delta(x)$, the corresponding power is simply given by $P_{\delta}(k) \propto |\delta(k)|^2$. We normalize the DPS \citep{zhan2005} as,
\begin{equation}
\sigma^2_{F_{\delta}} = \int \limits_{-\infty}^{\infty}  \frac{dk}{2 \pi} \; P_{\delta}(k)
\end{equation}
where $\sigma^2_{F_{\delta}}$ is variance of the 1D density field. We bin the DPS in $20$ equispaced logarithmic bins in the range $\log k =0.301$ to  $2.466$ with bin width of $\Delta \log k = 0.114$ \citep{kim2004}.  

Following \citet{rollinde} and \citet{gaikwad2017a}, we take the average of all DPS along different sightlines in a mock sample (consisting of $20$ lines of sight). We then calculate the mean DPS and the associated errors from the mock suite (which consists of $N = 100$ mock sample). Let $P_{\delta,n}(k_i)$ denotes the value of DPS in $i^{th}$ bin of $n^{th}$ mock sample, then the average DPS in $i^{th}$ bin is given by,
\begin{equation}{\label{eq:stat-avg}}
\begin{aligned}
\overline{P}_{\delta}(k_i) &= \frac{1}{N} \; \sum \limits_{n=1}^{N} P_{\delta,n}(k_i) \;\; .\\
\end{aligned}
\end{equation}
The covariance matrix element $C(i, j)$ between the $i^{th}$ and $j^{th}$ bins is given by,
\begin{equation}{\label{eq:stat-cov}}
\begin{aligned}
C(i,j) &= \frac{1}{N-1} \; \sum \limits_{n=1}^{N} [\overline{P}_{\delta}(k_i) - P_{\delta,n}(k_i)]  [\overline{P}_{\delta}(k_j) - P_{\delta,n}(k_j)] \\
\end{aligned}
\end{equation}
where, $i$ and $j$ can take values from 1 to the number of bins. The above analysis assumes a mock sample path length of 1000$h^{-1}$ cMpc (i.e., the mock sample consisting of $20$ spectra, corresponding dimensionless absorption path length is $X \sim 5.35$ at $z=3$). We have done the similar analysis for the $5000 h^{-1}$ cMpc mock sample path length (i.e., the mock sample consisting of $100$ spectra, corresponding to $X \sim 26.75$ at $z=3$). In this case we find that the covariance matrix elements are similar to those from mock samples with $1000h^{-1}$ cMpc path length for all the statistics (see the discussion in Appendix \ref{sec:SNR-effect}). Hereafter unless mention the results are presented for mock sample path length of $1000h^{-1}$ cMpc. 

The top left panel of Fig. \ref{fig:delta-ps-results} shows the DPS for the \gthree (black circles), \ltf (blue squares) and \htf (red stars) models at $z=2.5$. The grey shaded region is the $1\sigma$ uncertainty coming from sample variance (i.e., variation in DPS along different sightlines) in the \gthree DPS. The bottom left panel shows the residual fraction (hereafter residual for simplicity) between \ltf (blue squares) and \htf (red stars with errorbars) model with respect to \gthree model. The residual between \htf (or \ltf) and \gthree model is defined as follows,
\begin{equation}\label{eq:residual}
\mathcal{R}_{\rm \htf} = 1 - \frac{P_{\delta,{\rm \htf}}}{P_{\delta,{\rm GADGET-3}}} \;\;\;\; .
\end{equation}
The errors on \htf residuals in bottom left panel of Fig. \ref{fig:delta-ps-results} correspond to grey shaded region in top left panel (sample variance). Other panels in Fig. \ref{fig:delta-ps-results} are similar to left most panels but for different redshifts. The redshifts are mentioned in each panel.
 
 Although we show the DPS in the range $1 \leq k \leq 300 h \: {\rm Mpc}^{-1}$, all these scales are not accessible in the current set of best possible spectroscopic observations. The first scale is introduced by the typical temperature of the IGM at cosmic mean density $\sim 10^4$ K which corresponds to a velocity smoothing of $\sim 12$ \kmps. The second scale is spectral resolution of $\sim 7$ \kmps achieved by the current echelle spectrographs like HIRES or UVES. Thus scales below $\sim 7$ \kmps or the Fourier modes above the cut-off scale $k_{\rm cutoff}(z=3) = 101h \: {\rm Mpc}^{-1}$ ($k_{\rm cutoff}$ varies with redshift and is shown by magenta dashed vertical line) cannot be probed by \lya forest observations. Note that the velocity sampling of our simulated mock spectra is $\sim 1.11$ \kmps. 
 
We find that at all redshifts the DPS for \htf and \ltf are within 12 and 18 percent (and well within $1\sigma$ uncertainty due to sample variance) respectively with that from \gthree at $k \leq k_{\rm cutoff}$.  The \ltf model has higher power on the scales in the range 35-135 ckpc ($k \sim 180-47 \:h$ \mpc$^{-1}$). Note that although the instantaneous TDR is similar in \htf and \ltf model (see Fig. \ref{fig:eos-comparison}), the thermal history of the particles is different because the particles in \ltf model that do not go through shocks are effectively evolved at temperature smaller by factor of 100 as compared to those in \htf. The pressure smoothing scale, in addition to instantaneous TDR, also depends on thermal history of the particles \citep{girish2015}.  Thus the density field in \ltf model is less smooth (and hence has more power) as compared to that from \htf model at small scales. This difference is more prominent at high redshifts. This highlights the need for an appropriate smoothing of the density field on scales larger than pressure smoothing scale for \ltf at higher redshifts. However, when we evolve our simulations with a high temperature floor (i.e., $T \sim 10^4$ K), Jeans length based on a instantaneous $T$ and $\Delta$ is adequate to capture the pressure smoothing effects over the scales probed by the \lya forest observations.

We notice at $k  > k_{\rm cutoff}$, the power in \htf model is smaller as compared to \gthree model. This is due to the fact that the minimum temperature before applying {\sc cite} (irrespective of the density) in \htf model is $\sim ~ 10^4$ K. However in \gthree model, the particles with $\Delta < 1$ are at temperature smaller than $10^4$ K (see Fig. \ref{fig:eos-comparison}). Thus higher temperature for $\Delta < 1$ particles in \htf model leads to an additional pressure smoothing, thus the power on scales $k > k_{\rm cutoff}$ is smaller than that from \gthree. However it is important to note that (for reasons mentioned above) the mismatch between \htf and \gthree model at $k > k_{\rm cutoff}$ does not have a significant effect on the \lya flux statistics presented later.

\subsection{Flux probability distribution function (FPDF)} 
\label{subsec:flux-pdf}

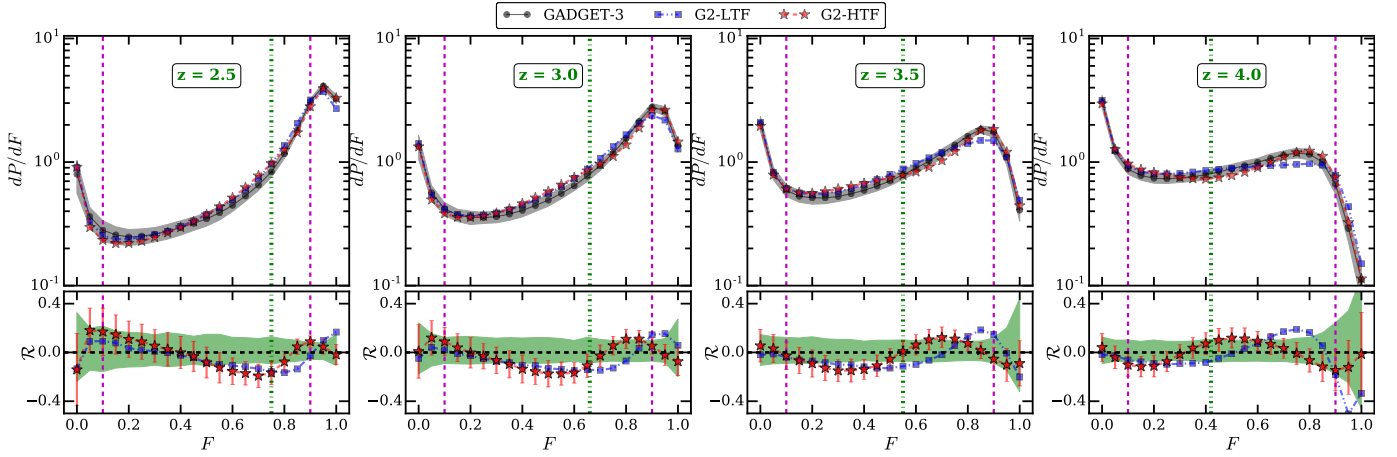
\InputFigCombine{Flux_PDF_z_All.pdf}{180}{Top left panel and bottom left panels are similar to the corresponding panels in Fig. \ref{fig:delta-ps-results} but for the FPDF statistics. Unlike Fig. \ref{fig:delta-ps-results}, the uncertainty in top (gray shaded region) and bottom (red stars with errorbars) left panel has contribution from the sample variance and finite noise added to the spectra.  The green shaded region in the bottom panels represents sample variance for \gthree model calculated using noise free spectra (i.e., SNR$=\infty$). The FPDF is compared in the range $0.1 \leq F \leq 0.9$ (shown by magenta dashed vertical lines) since the flux near continuum and saturated region is usually affected by observational systematics (see section \ref{subsec:flux-pdf} for details). The green dashed dot vertical line shows the mean flux for \gthree model. We present the results for $4$ redshifts whose values are mentioned in the corresponding panel. All the results are presented for SNR = 25, \GTW$=1$ and for mock sample path length of  $1000h^{-1}$ cMpc (corresponding to $X \sim 5.35$ at $z=3$).}{\label{fig:flux-pdf-results}}

The FPDF is one of the flux statistics that is relatively straightforward to calculate from observations as well as simulations \citep{jenkins1991,mcdonald2000,kim2007,vincent2007,rollinde,gaikwad2017a}. Note that we have added the noise to flux corresponding to the SNR of 25. Unless mentioned, hereafter all the results are presented for SNR=25 . We calculate the FPDF in $21$ equally spaced bins with bin centres in the range $F = 0.05$ to $1.0$ and bin width $\Delta F = 0.05$ \citep[consistent with][]{kim2007}. The pixels with $F < 0$ ($F > 1$) are included in the first (last) bin. Let $P_{n}(F_i)$ denote the value of FPDF in $i^{th}$ bin of $n^{th}$ mock sample then average FPDF in $i^{th}$ bin (denoted as $\overline{P}(F_i)$) is given by Eq. \ref{eq:stat-avg} where we replace $P_{\delta,n}(k_i)$ with $P_{n}(F_i)$. Similarly the covariance matrix element $C(i, j)$ between the $i^{th}$ and $j^{th}$ bins is obtained from Eq. \ref{eq:stat-cov} by replacing $P_{\delta,n}(k_i)$, $\overline{P}_{\rm \delta}(k_i)$ with $P_{n}(F_i)$,  $\overline{P}(F_i)$ respectively.

Fig. \ref{fig:flux-pdf-results} shows the comparison of FPDF obtained from \gthree, \htf and \ltf models. Various symbols and line styles are same as those used in Fig. \ref{fig:delta-ps-results}. Note that for all the three models we use $\Gamma_{12} =1$, SNR=$25$ and a path length of $1000h^{-1}$ cMpc  ($X \sim 5.35$ at $z=3$) for the mock sample. We calculated the mean FPDF, the covariance matrix and residuals ($\mathcal{R}$) using Eq. \ref{eq:stat-avg}, \ref{eq:stat-cov} and \ref{eq:residual} respectively with $P_{\rm \delta}$ replaced by $P_{\rm F}$.  The $1\sigma$ uncertainty in top (gray shaded region) and bottom left panels (red stars with errorbars) are contributed by the uncertainty in FPDF along different sightlines (sample variance) and finite SNR of the spectra. In order to separate out the statistical error arising purely from the assumed SNR and that from the sample variance, we calculate the FPDF for noise free spectra (i.e. SNR$=\infty$ but same $\Gamma_{12}$ and mock sample path length) for \gthree model. The green shaded region in bottom panels of Fig. \ref{fig:flux-pdf-results} represents sample variance from \gthree model. The size of this sample variance is comparable to the errors on FPDF from \gthree model (red stars with errorbars) suggesting that the errors are dominated by sample variance.  The mean flux in \gthree model (shown by green dashed dot vertical line) differs by less than 0.78 percent with that from \htf and \ltf models at all $z$.

The flux near the normalized continuum (in bins with $F>0.9$) is usually affected by two observational systematics, (i) the continuum fitting of the spectra\footnote{This systematic is severe at $z \geq 3.5$ as the continuum is not well defined and many pixels are near the saturated region (i.e., $F < 0.1$).} and (ii) the noise property of the spectra. On the other hand, flux near saturated region ($F<0.1$) depends on accurate background sky subtraction. Thus while comparing observed FPDF and the simulated FPDF, one usually compares them in the range $0.1 \leq F \leq 0.9$ \citep{gaikwad2017a}. In Fig. \ref{fig:flux-pdf-results}, we compare the FPDF from three models within the range $0.1 \leq F \leq 0.9$ shown by magenta dashed vertical lines.

It is clear from the bottom panels of Fig. \ref{fig:flux-pdf-results} that the difference between FPDF in the range $0.1 \leq F \leq 0.9$ for \htf and \ltf (with respect to \gthree model) is less than $15$ and $18$ percent respectively at all redshifts. Note that sample variance is typically of the order of 13 percent in the range $0.1 \leq F \leq 0.9$. 

\subsection{Flux power spectrum (FPS)} 
\label{subsec:flux-ps}
\InputFigCombine{Flux_PS_z_All.pdf}{180}{Comparison of FPS obtained from \gthree, \ltf  and \htf models. The symbols and line styles are same as in Fig. \ref{fig:flux-pdf-results}. The FPS from \ltf and \htf model is consistent within 5 percent with that from \gthree model (see section \ref{subsec:flux-ps} for details). The magenta dashed vertical line shows the wavelet scale used in section \ref{subsec:flux-wt}.}{\label{fig:flux-ps-results}}
Like the DPS, the FPS is a two point correlation function between pixels of the \lya transmitted flux \citep{croft1998,mcdonald2000,mcdonald2003,kim2004,zhan2005,prats2015}. The FPS is known to be sensitive to the astrophysical parameters such as \GHI, $T_0$ and $\gamma$ \citep{zaldarriaga2001,zaldarriaga2002,viel2004a} in addition to the cosmological parameters. The procedure for calculating the FPS is identical to that of the DPS. If we denote the value of FPS in $i^{th}$ bin of $n^{th}$ mock sample as $P_{F,n}(k_i)$ then the average FPS in $i^{th}$ bin is obtained from Eq. \ref{eq:stat-avg} by replacing $P_{\delta,n}(k_i)$ with $P_{F,n}(k_i)$. In similar vein, the covariance matrix elements $C(i,j)$ are obtained from Eq. \ref{eq:stat-cov}. The $\chi^2$ is calculated using the full covariance matrix.

In Fig. \ref{fig:flux-ps-results} we compare the FPS between different models. Note that at the redshift of interest the astrophysical parameters \GHI, $T_0$, $\gamma$ and spectral properties such as SNR, resolution and mock sample path lengths are same for different models. The FPS for different models behave in a way similar to the DPS. The FPS obtained from \htf model is consistent within $1\sigma$ and $5$ percent accuracy with that from the \gthree model at all redshifts. However, \ltf models at $z=3.5$ and $4.0$ have slightly excess power (but still within 5 percent) at scales in the range $20-100$ \kmps ($k \sim 0.06-0.28 \; {\rm s \: km}^{-1}$).  Similar to DPS, this excess power in the \ltf model can be attributed to the differences in the thermal history of the particles. We also see that the sample variance in FPS is smaller as compared to DPS as noted by \citet{zhan2005}. This is because the transformation (logarithmic suppression) between baryon density and flux is non-linear.
\vspace{-2mm}
\subsection{Wavelet statistics} 
\label{subsec:flux-wt}
\InputFigCombine{Flux_WT_z_All.pdf}{180}{Comparison of wavelet PDF obtained from \gthree, \ltf  and \htf models. The symbols and line styles are same as in Fig. \ref{fig:flux-pdf-results}. The wavelet PDF from \htf model is within 16 percent with that from \gthree model while the wavelet PDF from  \ltf is significantly different at $z \geq 3$ than \gthree model.}{\label{fig:wavelet-results}}

\begin{table}
\caption{Median wavelet power ($\log A_{L,n}$) in 4 different redshifts bins for the \gthree, \ltf and \htf models. The intervals represent 68 percentile around median.}
\begin{center}
\begin{tabular}{cccc}
\hline \hline
$z$	& \gthree & \ltf & \htf \\ \hline
2.5 & $-2.93 \pm 0.12$ & $-2.91 \pm 0.12$ & $-2.92 \pm 0.12$ \\ 
3.0 & $-2.91 \pm 0.12$ & $-2.87 \pm 0.12$ & $-2.91 \pm 0.12$ \\ 
3.5 & $-2.85 \pm 0.13$ & $-2.76 \pm 0.12$ & $-2.86 \pm 0.13$ \\ 
4.0 & $-2.76 \pm 0.15$ & $-2.57 \pm 0.13$ & $-2.77 \pm 0.14$ \\   \hline \hline
\end{tabular}
\end{center}
\label{tab:median-wavelet}
\end{table}

The wavelet statistic has been used in the past to constrain $T_0$ and $\gamma$ of the IGM \citep{theuns2000b,theuns2002,zaldarriaga2002,lidz2010,garzilli2012}. Wavelets have finite support in both real and Fourier space and thus can be used to extract the power at scales of interest. This is necessary because large scales (small $k$) are not sensitive to $T_0$, $\gamma$ variation whereas small scales (large $k$) are contaminated by noise and metal lines in observations \citep{lidz2010}. We use the ``Morlet'' wavelet, usually a sine (or a cosine) function damped by Gaussian, which has the form
\begin{equation}
\begin{aligned}
\Psi(x) = A \; \exp(-i \: k_0 \: x) \; \exp \bigg[ -\frac{x^2}{2 \: s^2_n}\bigg]
\end{aligned}
\end{equation}
where $s_n = 35$ \kmps,  and $k_0 = s_n / 2 \pi$ is the scale over which power is extracted. As shown by \citet{lidz2010}, this scale is sensitive to $T_0$ and $\gamma$ variations. $A$ is a normalization constant fixed by,
\begin{equation}
\begin{aligned}
\int \limits_{-\infty}^{\infty} |\Psi(x)|^2 \; dx = 1 \;\;\;\; .
\end{aligned}
\end{equation}
The wavelet coefficients are obtained by convolving the \lya flux ($F$) with Morlet as,
\begin{equation}
\begin{aligned}
a_n(x) = \int \limits_{-\infty}^{\infty} F(x^{\prime}) \; \Psi(x-x^{\prime}) \; dx^{\prime} 
\end{aligned}
\end{equation}
The wavelet power is then given by $A_n(x) = |a_n(x)|^2$. Following \citet{lidz2010}, we smooth the wavelet power on scales of $L=1000$ ${\rm km \; s}^{-1}$ to avoid noisy excursions in wavelet power
\begin{equation}
\begin{aligned}
A_{L,n}(x) = \frac{1}{L} \; \int \limits_{-\infty}^{\infty} \Theta(|x-x^{\prime}|; L/2) \; A_n(x^{\prime}) \; dx^{\prime} 
\end{aligned}
\end{equation}
where $\Theta(|x-x^{\prime}|; L/2)$ is the top-hat filter. It is important to note that wavelet power is anti-correlated with $T_0$ i.e., the wavelet power is smaller for higher $T_0$ and vice-versa.

Fig. \ref{fig:wavelet-results} shows the comparison of the PDF of the smoothed wavelet power $(A_{L,n})$ (hereafter wavelet PDF) from the three models. As the TDR parameters $T_0$ and $\gamma$ evolve with redshift, the peak and amplitude of the wavelet PDF also evolves accordingly. The bottom panels in Fig. \ref{fig:wavelet-results} show that the wavelet PDF for \htf model is within sample variance (green shaded region) and within 18 percent (red stars with errorbars) to that from the \gthree model at all redshifts. In contrast, the wavelet PDF is systematically shifted to larger values at higher redshifts for the \ltf model as compared to the \gthree model even though the thermal history parameters are quite similar at those redshifts (see Fig. \ref{fig:T0-gamma-evolution}). 
This can also be seen from Table. \ref{tab:median-wavelet} where the median wavelet power is consistently larger at higher redshift. Since the distribution is skewed, the errors given in Table. \ref{tab:median-wavelet} correspond to 68 percentile around the median value. The median wavelet power from \htf model is in good agreement (0.4 percent) with that from \gthree. However, median wavelet power in \ltf model is consistently lower than that from \gthree model at higher redshifts ($z \geq 3.5$, difference $\sim 7$ percent).  
This is because the wavelet scale used in our analysis ($s_{\rm n} = 35$ \kmps) corresponds to $k\sim 0.18 \; {\rm s \; km}^{-1}$ shown by magenta dashed vertical line in Fig. \ref{fig:flux-ps-results}. At this scale, the \ltf FPS has larger power as compared to \gthree FPS due to difference in density evolution and thermal history of the particles. Thus corresponding wavelet power is also larger for \ltf model as compared to \gthree model. Note that the wavelet power is still large in \ltf model even if we use a factor $\sim 2$ higher \GHI (corresponding to best fit value for \ltf model see section \ref{subsec:chisq-analysis}).

Due to such systematics, the inferred $T_0$ from \ltf model (or models in which Jeans smoothing effect from thermal history are not accounted for) would be larger at higher redshift and may lead to a misinterpretation of earlier He~{\sc ii} reionization. On the other hand \htf model though approximate in computing the Jeans smoothing does produce consistent results with that from \gthree. Thus the inferred $T_0$ from \htf model doesn't seem to be skewed by any systematic discussed above.
\InputFigCombine{Flux_CS_z_All.pdf}{180}{Comparison of curvature PDF obtained from \gthree, \ltf  and \htf models. The symbols and line styles are same as in Fig. \ref{fig:flux-pdf-results}. The curvature PDF from \htf model is within 10 percent with that from \gthree model at all redshifts. The sample variances are usually higher than these differences. In contrast, the curvature is systematically larger at higher redshifts in the \ltf than in the \gthree model (residual is as high as 120 percent).}{\label{fig:curvature-results}}
\subsection{Curvature statistics} 
\label{subsec:flux-cs}

\begin{table}
\caption{Median curvature in 4 different redshifts bins for the \gthree, \ltf and \htf models.}
\begin{center}
\begin{tabular}{cccc}
\hline \hline
$z$	& \gthree & \ltf & \htf \\ \hline
2.5 & $-3.35 \pm 0.57$ & $-3.35 \pm 0.58$ & $-3.35 \pm 0.58$ \\ 
3.0 & $-3.32 \pm 0.60$ & $-3.29 \pm 0.61$ & $-3.31 \pm 0.60$ \\ 
3.5 & $-3.29 \pm 0.63$ & $-3.22 \pm 0.65$ & $-3.27 \pm 0.64$ \\ 
4.0 & $-3.28 \pm 0.69$ & $-3.17 \pm 0.72$ & $-3.26 \pm 0.70$ \\  \hline \hline
\end{tabular}
\end{center}
\label{tab:median-curvature}
\end{table}

Similar to the wavelet analysis, \citet{becker2011} introduced a curvature statistics to measure the amount of small-scale structure in the \lya forest. The curvature $\kappa$ is defined as,
\begin{equation}
\begin{aligned}
\kappa \equiv \frac{F^{\prime \prime}}{[1 + (F^{\prime})^2]^{3/2}}
\end{aligned}
\end{equation}
where $F^{\prime}$, $F^{\prime \prime}$ is first and second derivative of \lya transmitted flux respectively. This statistics is suitable for obtaining the IGM temperature at characteristic overdensity which is found to be an almost one-to-one function of the mean curvature regardless of $\gamma$ \citep{becker2011,boera2014,hamsa2014,hamsa2015,sanderback2016}. Following the earlier works, \citet{hamsa2015} have shown that the mean and the percentiles of the curvature distribution function can be used to obtain constraints on the TDR. The comparisons of curvature PDF from three models are shown in Fig. \ref{fig:curvature-results}. The curvature PDF from \htf model is within 10 percent to that from \gthree model at all redshifts. On the other hand,  curvature is systematically more at higher redshifts for \ltf models than those of \gthree models and the residuals are as high as $\sim 120$ percent. The median curvature values are summarized in Table \ref{tab:median-curvature}. The median curvature in \ltf model is systematically higher than that from \gthree model at $z=4$. This is because of small scale fluctuations in density field (and hence flux) are larger for the \ltf model (see Fig. \ref{fig:flux-ps-results}) which affects the curvature measurement. Thus similar to wavelet analysis, the $T_0$ inferred from \ltf model using curvature statistics would be larger at higher redshifts ($z \geq 3.5$). Whereas $T_0$ inferred from \htf model would be consistent with that from \gthree model.

\subsection{Column Density Distribution function (CDDF)} 
The next three statistics we will discuss treat the \lya forest as a composition of discrete clouds. Each \lya line is fitted with multiple Voigt profiles each having 3 free parameters column density (${\rm N_{HI}}$), linewidth ($b$) parameter and line center ($\lambda_{\rm c}$). We used ``VoIgt profile Parameter Estimation Routine'' {\sc (viper)} to decompose the \lya forest  into multi-component Voigt profiles. More details can be found in \citep{gaikwad2017b}.

\InputFigCombine{CDD_z_All.pdf}{180}{Comparison of CDDF obtained from \gthree, \ltf  and \htf models. The symbols and line styles are as in Fig. \ref{fig:flux-pdf-results}. The incompleteness of the sample is not accounted for in the calculation of CDDF. The CDDF from \htf is consistent within $18$ percent with that from \gthree (red stars with errorbars).}{\label{fig:cdd-results}}

The CDDF, $f({\rm N_{HI}},z)$, is a bivariate distribution that describes the number of absorption lines with column density in range \logNHI to \logNHI + \dlogNHI and redshift in the range $z$ to $z + dz$. The CDDF is sensitive to \GHI \citep{schaye2001,shull2012b,kollmeier2014,shull2015,gaikwad2017b,viel2016,gurvich2017}. Fig. \ref{fig:cdd-results} shows the comparison of CDDF obtained from \gthree, \ltf and \htf models. 
We fit the noise free spectra from \gthree model and calculate the sample variance as shown by green shaded region in bottom panels. We have not accounted for the incompleteness of the sample in the calculation of redshift path length. This affects the shape of the CDDF at low \logNHI end. The CDDF from \htf models agree within sample variance ($1.25 \sigma$) and consistent (within 18 percent) with that from \gthree at all redshifts. For \logNHI$>13.5$, the \ltf model is consistent within 10 percent at $z \leq 3.5$ except at $z=4$ where the differences are large $\sim 40$ percent. In addition, as expected the \ltf model predicts more number of lines at lower column densities i.e., \logNHI$<13.5$. This is because the features arising from small scale density fluctuations of \ltf model (as seen in Fig. \ref{fig:los-comparison}) is identified and fitted by {\sc viper} as narrow lines with smaller column densities (for example see the region between $8-9$ cMpc in Fig. \ref{fig:los-comparison}). However, like other statistics the CDDF from \htf model is in good agreement with that from \gthree model.

\subsection{Linewidth ($b$ parameter) distribution function} 
\InputFigCombine{BPD_z_All.pdf}{180}{Comparison of $b$ parameter distribution obtained from \gthree, \ltf  and \htf models. The symbols and line styles are as in Fig. \ref{fig:flux-pdf-results}. Each panel is similar to that from Fig. \ref{fig:flux-pdf-results} except the comparison is shown for $b$ parameter distrbution from \gthree (black circle), \ltf (blue squares) and \htf (red stars) models. The $b$ parameter distribution and residuals are plotted from all the lines in sample with relative error in $b$ parameter less than 0.3 (the lines below completeness limit are also included). The green shaded region in the bottom panels show the sample variance on $b$ parameter distribution from \gthree model. The $b$ parameter distribution is within $\sim 18$ percent agreement (at $b < 60$ \kmps) with that from the \gthree.}{\label{fig:bpd-results}}

\begin{table}
\caption{Median $b$ parameter (with 68 percentile interval) in 4 different redshifts bins for the \gthree, \ltf and \htf models.}
\begin{center}
\begin{tabular}{cccc}
\hline \hline
$z$	& \gthree & \ltf & \htf \\ \hline
2.5 & 25.03$^{+29.51}_{-9.54}$ & 22.53$^{+23.46}_{-9.80}$ & 22.76$^{+24.44}_{-8.84}$ \\ \\
3.0 & 26.61$^{+30.14}_{-10.02}$ & 23.22$^{+22.12}_{-9.65}$ & 24.69$^{+27.01}_{-9.58}$ \\ \\
3.5 & 30.15$^{+34.69}_{-11.96}$ & 25.69$^{+25.23}_{-10.62}$ & 28.86$^{+32.21}_{-11.54}$ \\ \\ 
4.0 & 35.43$^{+37.28}_{-15.13}$ & 29.22$^{+28.12}_{-12.83}$ & 34.21$^{+36.47}_{-14.51}$ \\  \hline \hline
\end{tabular}
\end{center}
\label{tab:median-b-parameter}
\end{table}

The top panels in Fig. \ref{fig:bpd-results} show the linewidth distribution, which is sensitive to thermal history, pressure smoothing and unknown turbulent motions in the IGM \citep{schaye1999,schaye2000,mcdonald2001,dave2001b,gaikwad2017b,viel2016},  from \gthree (black circles), \ltf (blue squares) and \htf (red stars) models for different redshift bins. The $b$ parameter distribution and residuals plotted in Fig. \ref{fig:bpd-results} is calculated from all the lines in sample with relative error in $b$ parameter smaller than 0.5.  Again, unlike the \ltf model the linewidth distribution from \htf model is consistent (within $18$ percent uncertainty) with that from \gthree model at $b < 60$ \kmps.  On the other hand residual between \ltf and \gthree model is large. 
It is interesting to note that the peak of the $b$ distribution shifts towards larger $b$ values with increasing redshift. This is because the lines tend to be saturated and blended together at higher redshifts. As a result, the fitted $b$ parameter tends to be larger. However, the errors on the fitted $b$ parameters are relatively higher at higher redshifts and hence it is non-trivial to constrain the thermal history of IGM at higher redshifts ($z > 3.5$) using line fitting method \citep{webb1991,fernandez1996}.
This can also be seen from Table \ref{tab:median-b-parameter} where we have summarize the median $b$ parameters for the $3$ models at 4 different redshifts. The median $b$ increases from $z = 2.5$ to $z=4.0$ in all models. The 68 percentile intervals around median are asymmetric because the $b$ distribution is skewed. Within $1\sigma$ errorbars the $b$ distribution from \ltf and \htf model is consistent with that from \gthree model.

\subsection{$b$ versus \logNHI scatter} 
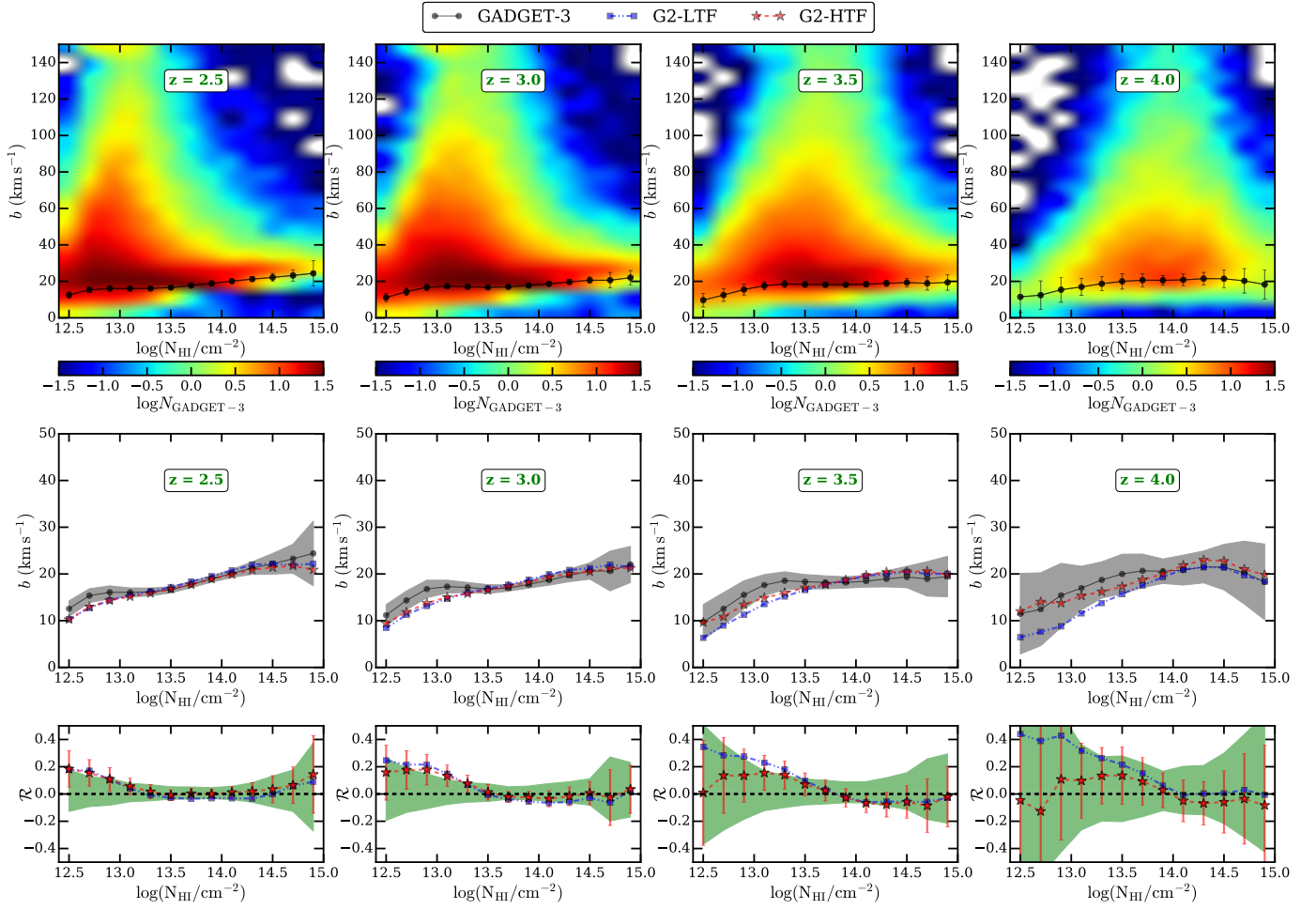
\InputFigCombine{b_vs_NHI.pdf}{180}{Top panels show the comparison of the $b$ vs \logNHI correlation from \gthree. The colour scheme indicates density of points in logarithmic units for \gthree. Middle and bottom panels show comparison of the $b$ vs \logNHI lower envelope for \gthree, \ltf and \htf models. The symbols and line styles are as in Fig. \ref{fig:flux-pdf-results}. The lower envelopes are obtained by calculating 10$^{\rm th}$ percentile of $b$ values in \logNHI bin. The lower envelopes for the three model are also shown in top panel.}{\label{fig:b-NHI-results}}

The top panels in Fig. \ref{fig:b-NHI-results} show the $b$ versus \logNHI scatter for \gthree model. The color scheme represents density of points in logarithmic units.  One way to assess the goodness of fit is to match the lower-envelope in $b$ versus \logNHI plot. The lower-envelope in the $b$ versus \logNHI plot has been used in the past to constrain the thermal history parameters $T_0$ and $\gamma$ \citep{schaye1999,schaye2000,mcdonald2001}. Following \citet{garzilli2015a}, we obtain the lower-envelope by calculating the $10^{\rm th}$ percentile of $b$ values in \logNHI bin. The middle and bottom panels show the comparison of the lower-envelope obtained from \gthree, \htf and \ltf model. The lower envelope in $b$ versus \logNHI plot from \htf 
(red stars) is within sample variance and in 18 percent agreement with \gthree (black circles). On the other hand, at $z=3.5$ and $4.0$, the lower-envelope from \ltf (blue stars) is consistently smaller at \logNHI$<13.5$ than that from \gthree model. This can again be attributed to extra absorption line features with smaller \logNHI identified by the {\sc viper}. 
Because of the systematically smaller lower-envelop in \ltf model at \logNHI$<13.5$, $\gamma$ derived from \ltf model would be systematically larger than that from \gthree model at $z \geq 3.5$.

To summarize the results presented in sections 4.1--4.8, we find that the \lya statistics derived from \htf model are within 20 percent (except for $b$ parameter distribution) to that from \gthree model and within the sample variance for a path length of $\sim 1000 h^{-1}$ cMpc ($X \sim 5.35$ at $z=3$).  
\subsection{$\chi^2$ analysis}
\label{subsec:chisq-analysis}
The main motivation of this work is to develop the method to simulate the \lya forest in order to efficiently explore the parameter space. Hence, it is important to show the accuracy of the method in recovering the astrophysical parameters. In this section we present the $\chi^2$ analysis and show the accuracy of our method in recovering the \HI photoionization rate \GHI.

The differences we see between the FPDF and the FPS from different models will have direct consequence in the derived parameter values like \GHI. To study this, we treat \gthree as the reference model and see how the value of \GHI is recovered when we use the \ltf and \htf models. Note that we use the noise (corresponding to SNR=25) added  \lya transmitted flux in all the models. We vary \GTW in \htf (or \ltf) model and calculate the FPDF and FPS. The $\chi^2$ between the FPDF / PS calculated from \gthree and that from \htf (or \ltf) model can be written in the matrix form as \citep[for similar method see][]{gaikwad2017a},
\begin{equation}
\begin{aligned}
\chi^2_{(\Gamma_{\rm 12})} = [P_{(\Gamma_{\rm 12})} - P_{\rm fid}]  \; C^{-1} \; [P_{(\Gamma_{\rm 12})} - P_{\rm fid}]^{\rm T}
\end{aligned}
\end{equation}
where $P_{\rm fid}$ and $P_{(\Gamma_{\rm 12})}$ is flux statistics (either FPDF or PS) from \gthree and \htf (or \ltf) model respectively. $C$ is the covariance matrix as given in Eq. \ref{eq:stat-cov}. Note that we use full covariance matrix for $\chi^2$ estimation. \\
\textbf{\GTW recovery:} 
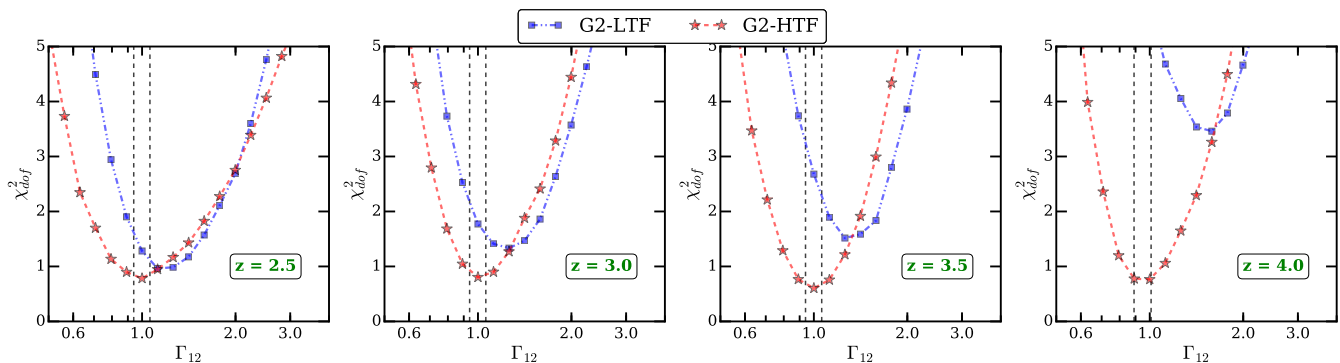
\InputFigCombine{Gamma_12_constraints_New.pdf}{175}{Different panels show the recovery of \GHI for different redshift (given in each panel) using FPDF and FPS stastics. The combined (for FPDF and PS)  reduced $\chi^2$ as a function of \GTW for \ltf (blue squares) and \htf (red stars) model is shown in each panel. \gthree is used as the reference model with \GTW = 1. The $\chi^2$ is calculated between statistics from \gthree and \ltf or \htf models (see Table \ref{tab:chi-square-values}). The $1\sigma$ statistical uncertainty on the recovered \GTW for \htf model is indicated by black dashed vertical lines. The above analysis is done for SNR = 25 and for mock sample path length of  $1000h^{-1}$ cMpc  (corresponding to $X \sim 5.35$ at $z=3$).}{\label{fig:Gamma12-recovery}}
The panels in Fig. \ref{fig:Gamma12-recovery} show reduced $\chi^2$ as a function of \GTW from \htf (red stars) and \ltf (blue squares) model for four different redshifts. The black dashed vertical lines show the statistical uncertainty in \GTW for \htf model\footnote{Under the assumption of normal distribution, the statistical uncertainty corresponds to $\chi^2 = \chi^2_{\rm min} + \Delta \chi^2$ where $\Delta \chi^2 = 1$ \citep{press1992}.}. The \GTW is recovered within $1\sigma$ ($d$\GTW$\sim \pm 0.05$, within $5$ percent accuracy) in \htf model at all redshifts whereas \ltf model fails to recover the \GTW within $1\sigma$. The \GTW recovered from \ltf model at $z=3.0$ and $z=4.0$ is higher by a factor of  1.7 and 2 respectively. The minimum $\chi^2_{\rm dof}$ for \htf model is also close to 1 indicating the goodness-of-fit. Note that the above analysis is done assuming  HM12 UVB, SNR = 25 and mock sample path length of  $1000h^{-1}$ cMpc (corresponding $X \sim 5.35$ at $z=3$). However, the effect of SNR on \GHI recovery is not significant. To illustrate this, we recover the \GHI at four different redshift bins for SNR varying from 15 to $\infty$ (noise free spectra) as shown in Table \ref{tab:SNR-effect} of Appendix \ref{sec:SNR-effect}.  We can see that for wide range of SNR, the \GHI is recovered in the case of \htf model within $5$ percent accuracy. The statistical uncertainty on recovered \GHI is also very similar\footnote{The $\chi^2_{\rm dof}$ increases by $\sim 60$ percent for SNR=$\infty$ as compared to SNR$=15$ due to smaller errorbars in earlier.}. This is because, as pointed out earlier, the scatter in FPDF and FPS mainly comes from sample variance. Thus effect of SNR is not important in recovery of \GHI. The main conclusion from this study is that the gas pressure is very important while deriving \GHI based on FPDF and FPS. Simple pressure smoothing based on instantaneous temperature and density values can lead to an overestimation of \GHI. We show that if we run \gtwo with higher temperature floor ($T \sim 10^4$ K), we are able to recover the correct \GHI within $5$ percent accuracy using our approach. \\
\textbf{$\chi^2$ analysis for other statistics:} We have also calculated the $\chi^2$ between \htf and \ltf model with \gthree as a reference model keeping all other parameters fixed. Table \ref{tab:chi-square-values} summarizes the reduced $\chi^2$ values for different statistics. Note that we calculate the reduced $\chi^2$ values for the statistics other than FPDF and FPS using only the diagonal terms of the covariance matrix as the off-diagonal terms are noisy. It is clear from Table \ref{tab:chi-square-values} that the reduced $\chi^2$ is generally  less for \htf model as compared to that for \ltf model which suggest \htf model is in general better agreement with \gthree model. The $\chi^2$ analysis for the case of an enhanced UVB is presented in Appendix \ref{sec:UVB-effect} where we show that the \htf model is consistent with \gthree model for a significantly different thermal history.
\begin{table*}
\caption{Reduced $\chi^2$ between \ltf, \htf model and reference model \gthree for different statistics.}
\begin{threeparttable}
\centering
\begin{tabular}{lcccccccc}
\hline \hline
& \multicolumn{2}{c}{$z=2.5$} & \multicolumn{ 2}{c}{$z=3.0$} & \multicolumn{ 2}{c}{$z=3.5$} & \multicolumn{ 2}{c}{$z=4.0$} \\ 
Statistics\tnote{1} & \ltf & \htf & \ltf & \htf & \ltf & \htf & \ltf & \htf \\ 
\hline 
Density Power spectrum ($\delta$) & 0.41 & 0.98 & 0.67 & 0.79 & 0.82 & 0.60 & 1.21 & 0.58 \\ 
FPDF & 1.26 & 0.76 & 1.41 & 0.80 & 1.79 & 0.76 & 1.87 & 0.81 \\ 
FPS & 0.50 & 0.36 & 0.68 & 0.25 & 1.82 & 0.22 & 3.23 & 0.45 \\ 
Wavelet PDF & 0.20 & 0.42 & 0.60 & 0.17 & 4.46 & 0.27 & 12.84 & 0.74 \\ 
Curvature PDF & 0.30 & 0.19 & 1.34 & 0.49 & 6.81 & 0.82 & 17.07 & 0.78 \\ 
CDDF & 2.84 & 1.20 & 2.85 & 1.07 & 4.51 & 0.47 & 4.48 & 0.41 \\ 
$b$ parameter distribution & 4.38 & 1.04 & 6.09 & 0.75 & 4.56 & 0.42 & 2.18 & 0.73 \\ 
$b$ vs \logNHI correlation & 0.70 & 0.54 & 1.27 & 0.46 & 1.35 & 0.44 & 1.55 & 0.62 \\ 
\hline \hline
\end{tabular}
\begin{tablenotes}
\item[1] For a given redshift, all the astrophysical parameters ($T_0$, $\gamma$, \GHI) are same for \ltf, \htf and \gthree models. Reduced $\chi^2$ is calculated using full covariance matrix for FPDF and FPS. However, for other statistics we used diagonal elements of the covariance matrix as off diagonal elements are noisy.
\end{tablenotes}
\end{threeparttable}
\label{tab:chi-square-values}
\end{table*} \\
\subsection{Effect of different thermal history}
The comparison between different models discussed in sections 4.1-4.9 has been performed for the ionization and heating rates from HM12 UVB model. It is, however, important to validate our method for different UVB models where the thermal history is significantly different from that in the case of the HM12 UVB. In Figs. \ref{fig:delta-ps-results-HM12-enhanced}-\ref{fig:b-NHI-results-HM12-enhanced}, we validate our method (for \htf model) for a UVB models in which $T_0$ is increased by a factor of $\sim 2$ while $\gamma$ remains same at all redshifts (see Appendix \ref{sec:UVB-effect} for details). We find that the statistics derived from the \htf model is again consistent within 20 percent to that from \gthree model for such a different thermal history. Thus for a range of physically motivated  photo-heating rates from UVB calculations \citep[such as][]{khaire2015b,khaire2015a}, we can easily probe the $T_0-\gamma$ parameter space and calculate \lya flux in \htf model without performing full \gthree simulation. Therefore our method can be a good first step to narrow down the parameter space before confirming the best fit parameter with \gthree simulation.
\subsection{Computational gain} 
We now highlight the advantages of using our method for simulating \lya forest:
\begin{itemize}
\item {\bf Efficiency :} Table \ref{tab:cpu-time} summarizes the CPU time consumption in various parts of the code. Significant fraction of time is spent in evolution of $\Delta$, $v$ and $T$ in both codes. However, unlike \gthree we need to evolve $\Delta$, $v$ and $T$ in \htf (or \ltf) only once. To vary astrophysical parameters in \htf, we just need to vary UVB in {\sc cite}. This allows one to probe $T_0$ and $\gamma$ parameter space efficiently. For example, the time (per core) required to simulate \lya forest for 10 different UVB in \gthree is $\sim 67$ days whereas for \gtwo is $\sim 8$ days.
\item {\bf Accuracy :} We have shown that our method (running \gtwo with a high temperature floor and post-process using {\sc cite}) produces statistical distributions that are consistent within sample variance (calculated from mock sample path length of $1000 h^{-1}$ cMpc or $X \sim 5.35$ at $z=3$) and within 20 percent to those obtained with \gthree. In particular, our method is accurate within 5 percent with that from \gthree in recovering \GHI.
\item {\bf Flexibility :} In addition to HM12, it is straightforward to incorporate other UVB such as \citet{faucher2009,khaire2015b,khaire2015a} in {\sc cite} and evolve the temperature without performing full hydrodynamic simulation. {\sc cite} can be run in either equilibrium or non-equilibrium ionization evolution mode. It is easy to incorporate cooling due to metals in {\sc cite} by changing cooling rate tables \citep[][for similar analysis]{wiersma2009,gaikwad2017a}\footnote{http://www.strw.leidenuniv.nl/WSS08/}.
\end{itemize}

Thus our method (though approximate) is efficient, flexible and sufficiently accurate to explore a large parameter space which otherwise would be more time consuming with self-consistent simulations like \gthree. However in practice, while constraining astrophysical parameters from observations, we propose to use the method in 3 steps (i) use \htf model to explore large parameter space and obtain the best fit parameters with corresponding statistical uncertainty, (ii) run \gthree simulation with best fit parameters (and also for parameters with $1\sigma$ deviation) and (iii) check if the statistics derived from data are consistent with those derived from \gthree model with best fit parameters.

Thus our method, while not a substitute for the full hydrodynamical simulation like \gthree, provide an efficient and reasonable accurate tool to explore a large parameter space which otherwise require large resources and computational time. A possible way to make use this method would be to narrow down the parameter space in the first step before confirming the best fit parameters with a full \gthree simulation.
\begin{table}
\caption{Consumption of CPU time (in hours) per core for various tasks of the code for a cosmological run from $z=99$ to $z=2.0$}
\begin{threeparttable}
\centering
\begin{tabular}{clcc}
\hline \hline
Step & Description\tnote{a} & \gthree & \htf \\ 
\hline 
1 & $\Delta$, $v$ and $T$ Evolution & 156 & 108  \\ 
2 & {\sc cite} ($T$ Evolution)\tnote{b} & $-$ & 3.5  \\ 
3 & Grid calculation\tnote{c} & 3.5  & 4  \\ 
4 & {\sc glass}\tnote{d} & 1  & 1  \\  
\hline 
-  & Total time to run  & 1605 & 193  \\ 
& 10 UVB model\tnote{e} &  ($67$ days) & ($8$ days) \\ 
\hline 
\end{tabular}
\begin{tablenotes}
\item[a] The analysis is done using 256 core on IUCAA PERSEUS cluster.
\item[b] {\sc cite} evolves the temperature of the SPH particles from $z=6.0$ to $z=2.0$. Temperature is evolved internally in \gthree.
\item[c] We used modified smoothing kernel for \htf or \ltf as given in Eq. \ref{eq:smth-kenrel-cases}. The time is given for 10240 random sightlines through simulation box.
\item[d] We apply TDR as given in Eq. \ref{eq:grid-eos} for \htf and \ltf models. The numbers are given for total 10 $\times$ 2048  simulated \lya forest spectra. We splice 5 sightline to cover redshift path length for a single spectra.
\item[e]  The total time required to run 10 UVB model for \gthree is sum of time consumed by steps 1, 3 and 4 (i.e. 160.5 $\times$ 10 hours). Unlike \gthree, step 1 is performed only once for \htf or \ltf models. For different UVB models, we follow step 2-4 in the post-processing stage. Hence the total time required to run 10 UVB model for \ltf or \htf model is (108 hours + 8.5 hours $\times$ 10 = 193 hours). 
\end{tablenotes}
\end{threeparttable}
\label{tab:cpu-time}
\end{table}
\vspace{-7mm}
\section{Summary}
\label{sec:summary}
With the advent of high quality observations, an efficient method to simulate the \lya forest would be useful for parameter estimation. Current state-of-art simulations like \gthree, though reproduce observational properties of  \lya forest very well, are computationally expensive for large parameter space exploration. As part of our ongoing effort, we have developed a post processing module for \gtwo called  ``Code for Ionization and Temperature Evolution'' ({\sc CITE}). In  \citet{gaikwad2017a}, we have shown that the predictions of our low  redshift simulations match well with other existing hydrodynamical simulations and estimated \GHI at $z < 0.5$ and  associated uncertainties using extensive exploration of the parameter space.  

For the resolution used in the above study  (gas particle mass $\delta m = 1.26 \times 10^7 \: h^{-1} \: {\rm M}_{\sun}$, pixel size $\delta x \sim 48.8h^{-1}$ ckpc), the pressure smoothing of baryons may not be a major issue.  However, for studying the high-$z$ \lya forest one usually uses higher resolution echelle data. When we use appropriate high resolution (gas particle mass $\delta m = 1.01 \times 10^5 \: h^{-1} \: {\rm M}_{\sun}$, pixel size $\delta x \sim 9.77h^{-1}$ ckpc) simulation boxes, we notice that  the density ($\Delta$) and velocity ($v$) fields are smoother for \gthree as compared to those from \gtwo. This is because the temperature and ionization state of the SPH particles in \gtwo is not calculated self-consistently (photo-heating and radiative cooling terms are not accounted for). In this work we show that by running a \gtwo simulation with elevated temperature floor  (i.e., $T \sim 10^4$ K) and using local Jeans smoothing we are able to appreciably overcome the above mentioned shortcomings of our method in the high resolution simulations.

The basic idea is to apply additional smoothing in \gtwo by a local Jeans length at the epoch of our interest. However, it is well known that  the smoothing in \gthree is not only decided by the instantaneous density and temperature of the particles but also to some extent by the thermal history of the particles. To understand this, we perform three high resolution simulations (gas particle mass $\delta m = 1.01 \times 10^5 \: h^{-1} \: {\rm M}_{\sun}$, pixel size $\delta x \sim 9.77h^{-1}$ ckpc) with same initial conditions  (i) \ltf : \gtwo with low temperature ($T \sim 100$ K) floor in which local Jeans length is decided by instantaneous density and temperature and (ii) \htf : \gtwo with high temperature ($T \sim 10^4$ K) floor in which even the unshocked gas is evolved at with a pressure appropriate for a photoionized gas at $T = 10^4$ K  and (iii) \gthree : a reference model for comparison with \ltf and \htf model. 

For \ltf and \htf models, we first estimate the temperature of SPH particles in \gtwo using our code {\sc cite}. We modify the smoothing kernel  to account for pressure smoothing and estimated the density, velocity field on grids. We find that the line of sight density and velocity from our method matches well with that from \gthree. We then compare our method for \ltf and \htf models with that from \gthree simulation. The main results of our analysis are as follows:  
\begin{itemize}
\item We obtain the evolution of thermal history parameters $T_0$ and $\gamma$ by estimating the temperature of the SPH particles from {\sc cite}. We show that the redshift evolution of  $T_0$ and $\gamma$ from \htf and \ltf are in very good agreement with that from \gthree. {\sc cite} also provides us with enough flexibility to solve the non-equilibrium ionization evolution equation. The $T_0$ and $\gamma$ evolution for non-equilibrium case is considerably different ($T_0$ is larger by $\sim 60$ percent and $\gamma$ is smaller by 15 percent at $z=3.7$) than that for equilibrium case. We show that the redshift evolution of $T_0$ and $\gamma$ for non-equilibrium case from our method is consistent with that from \citet[][difference less than 2.5 percent]{puchwein2015}.

\item We generate the \lya forest spectra by shooting random sightlines through simulation box in all the 3 models. The resulting \lya forest spectra along sightline are remarkably similar in the \htf and \gthree methods. However the \lya forest spectra in \ltf model show more variation as compared to that from \gthree. We compare the \ltf and \htf with the \gthree model using 8 different statistics, namely: (i) 1D density field PS,  (ii) FPDF, (iii) FPS, (iv) wavelet PDF, (v) curvature PDF, (vi) column density distribution function, (vii) linewidth distribution and (viii) $b$ vs \logNHI correlation, at four different redshift $z=2.5,3.0,3.5$ and $4.0$. Treating the \gthree model as the reference, we demonstrate that the \HI photoionization rate (\GHI) can be recovered, using FPDF and FPS statistics, well within $1\sigma$ statistical uncertainty using the \htf model. We find that the \htf model is in general very good agreement (within 20 percent and within $1\sigma$ sample variance calculated from $1000h^{-1}$ cMpc or $X \sim 5.35$ at $z=3$) with \gthree model at all redshifts. On the other hand \ltf model overestimates the \GHI by a factor of $\sim 2$ at  $z \geq 3.5$. 
 
\item Using enhanced HM12 photo-heating rates, we obtain a thermal history such that $T_0$ is increased by a factor of $\sim 2$. We show that our method for such significantly different thermal history is also consistent (in $1\sigma$) with \gthree simulation.

\end{itemize}

Our method to simulate the \lya forest is computationally less expensive, flexible to incorporate changes in UVB, metallicity, non-equilibrium ionization evolution etc. and accurate (in recovering \GHI) to within 5 percent. This method can be used in future more effectively to explore $T_0$, $\gamma$ and \GHI parameter space and to simultaneously constrain these quantities from observations.


\vspace{-5mm}
\section*{Acknowledgement}
All the computations are performed using the PERSEUS cluster at IUCAA and the HPC cluster at NCRA. We like to thank Volker Springel, Aseem Paranjape and Ewald Puchwein for useful discussion. We also thank the anonymous referee for improving this work and the manuscript.


\vspace{-5mm}
\bibliographystyle{mnras}
\bibliography{FLAG} 

\begin{thebibliography}{}
\makeatletter
\relax
\def\mn@urlcharsother{\let\do\@makeother \do\$\do\&\do\#\do\^\do\_\do\%\do\~}
\def\mn@doi{\begingroup\mn@urlcharsother \@ifnextchar [ {\mn@doi@}
  {\mn@doi@[]}}
\def\mn@doi@[#1]#2{\def\@tempa{#1}\ifx\@tempa\@empty \href
  {http://dx.doi.org/#2} {doi:#2}\else \href {http://dx.doi.org/#2} {#1}\fi
  \endgroup}
\def\mn@eprint#1#2{\mn@eprint@#1:#2::\@nil}
\def\mn@eprint@arXiv#1{\href {http://arxiv.org/abs/#1} {{\tt arXiv:#1}}}
\def\mn@eprint@dblp#1{\href {http://dblp.uni-trier.de/rec/bibtex/#1.xml}
  {dblp:#1}}
\def\mn@eprint@#1:#2:#3:#4\@nil{\def\@tempa {#1}\def\@tempb {#2}\def\@tempc
  {#3}\ifx \@tempc \@empty \let \@tempc \@tempb \let \@tempb \@tempa \fi \ifx
  \@tempb \@empty \def\@tempb {arXiv}\fi \@ifundefined
  {mn@eprint@\@tempb}{\@tempb:\@tempc}{\expandafter \expandafter \csname
  mn@eprint@\@tempb\endcsname \expandafter{\@tempc}}}

\bibitem[\protect\citeauthoryear{{Abramowitz} \& {Stegun}}{{Abramowitz} \&
  {Stegun}}{1972}]{abramowitz1972}
{Abramowitz} M.,  {Stegun} I.~A.,  1972, {Handbook of Mathematical Functions}

\bibitem[\protect\citeauthoryear{{Arinyo-i-Prats}, {Miralda-Escud{\'e}}, {Viel}
   \& {Cen}}{{Arinyo-i-Prats} et~al.}{2015}]{prats2015}
{Arinyo-i-Prats} A.,  {Miralda-Escud{\'e}} J.,  {Viel} M.,   {Cen} R.,  2015,
  \mn@doi [\jcap] {10.1088/1475-7516/2015/12/017}, \href
  {http://adsabs.harvard.edu/abs/2015JCAP...12..017A} {12, 017}

\bibitem[\protect\citeauthoryear{{Bahcall} \& {Peebles}}{{Bahcall} \&
  {Peebles}}{1969}]{bahcall1969}
{Bahcall} J.~N.,  {Peebles} P.~J.~E.,  1969, \mn@doi [\apjl] {10.1086/180337},
  \href {http://adsabs.harvard.edu/abs/1969ApJ...156L...7B} {156, L7}

\bibitem[\protect\citeauthoryear{{Becker} \& {Bolton}}{{Becker} \&
  {Bolton}}{2013}]{becker2013}
{Becker} G.~D.,  {Bolton} J.~S.,  2013, \mn@doi [\mnras]
  {10.1093/mnras/stt1610}, \href
  {http://adsabs.harvard.edu/abs/2013MNRAS.436.1023B} {436, 1023}

\bibitem[\protect\citeauthoryear{{Becker}, {Bolton}, {Haehnelt}  \&
  {Sargent}}{{Becker} et~al.}{2011}]{becker2011}
{Becker} G.~D.,  {Bolton} J.~S.,  {Haehnelt} M.~G.,   {Sargent} W.~L.~W.,
  2011, \mn@doi [\mnras] {10.1111/j.1365-2966.2010.17507.x}, \href
  {http://adsabs.harvard.edu/abs/2011MNRAS.410.1096B} {410, 1096}

\bibitem[\protect\citeauthoryear{{Bi} \& {Davidsen}}{{Bi} \&
  {Davidsen}}{1997}]{bi1997}
{Bi} H.,  {Davidsen} A.~F.,  1997, \mn@doi [\apj] {10.1086/303908}, \href
  {http://adsabs.harvard.edu/abs/1997ApJ...479..523B} {479, 523}

\bibitem[\protect\citeauthoryear{{Bi}, {Boerner}  \& {Chu}}{{Bi}
  et~al.}{1992}]{bi1992}
{Bi} H.~G.,  {Boerner} G.,   {Chu} Y.,  1992, \aap, \href
  {http://adsabs.harvard.edu/abs/1992A%26A...266....1B} {266, 1}

\bibitem[\protect\citeauthoryear{{Boera}, {Murphy}, {Becker}  \&
  {Bolton}}{{Boera} et~al.}{2014}]{boera2014}
{Boera} E.,  {Murphy} M.~T.,  {Becker} G.~D.,   {Bolton} J.~S.,  2014, \mn@doi
  [\mnras] {10.1093/mnras/stu660}, \href
  {http://adsabs.harvard.edu/abs/2014MNRAS.441.1916B} {441, 1916}

\bibitem[\protect\citeauthoryear{{Bolton} \& {Haehnelt}}{{Bolton} \&
  {Haehnelt}}{2007}]{bolton2007}
{Bolton} J.~S.,  {Haehnelt} M.~G.,  2007, \mn@doi [\mnras]
  {10.1111/j.1365-2966.2007.12372.x}, \href
  {http://adsabs.harvard.edu/abs/2007MNRAS.382..325B} {382, 325}

\bibitem[\protect\citeauthoryear{{Bolton}, {Haehnelt}, {Viel}  \&
  {Springel}}{{Bolton} et~al.}{2005}]{bolton2005}
{Bolton} J.~S.,  {Haehnelt} M.~G.,  {Viel} M.,   {Springel} V.,  2005, \mn@doi
  [\mnras] {10.1111/j.1365-2966.2005.08704.x}, \href
  {http://adsabs.harvard.edu/abs/2005MNRAS.357.1178B} {357, 1178}

\bibitem[\protect\citeauthoryear{{Bolton}, {Haehnelt}, {Viel}  \&
  {Carswell}}{{Bolton} et~al.}{2006}]{bolton2006}
{Bolton} J.~S.,  {Haehnelt} M.~G.,  {Viel} M.,   {Carswell} R.~F.,  2006,
  \mn@doi [\mnras] {10.1111/j.1365-2966.2006.09927.x}, \href
  {http://adsabs.harvard.edu/abs/2006MNRAS.366.1378B} {366, 1378}

\bibitem[\protect\citeauthoryear{{Bryan} et~al.,}{{Bryan}
  et~al.}{2014}]{enzo2014}
{Bryan} G.~L.,  et~al., 2014, \mn@doi [\apjs] {10.1088/0067-0049/211/2/19},
  \href {http://adsabs.harvard.edu/abs/2014ApJS..211...19B} {211, 19}

\bibitem[\protect\citeauthoryear{{Cen}, {Miralda-Escud{\'e}}, {Ostriker}  \&
  {Rauch}}{{Cen} et~al.}{1994}]{cen1994}
{Cen} R.,  {Miralda-Escud{\'e}} J.,  {Ostriker} J.~P.,   {Rauch} M.,  1994,
  \mn@doi [\apjl] {10.1086/187670}, \href
  {http://adsabs.harvard.edu/abs/1994ApJ...437L...9C} {437, L9}

\bibitem[\protect\citeauthoryear{{Choudhury}, {Srianand}  \&
  {Padmanabhan}}{{Choudhury} et~al.}{2001}]{trc2001}
{Choudhury} T.~R.,  {Srianand} R.,   {Padmanabhan} T.,  2001, \mn@doi [\apj]
  {10.1086/322327}, \href {http://adsabs.harvard.edu/abs/2001ApJ...559...29C}
  {559, 29}

\bibitem[\protect\citeauthoryear{{Cooke}, {Espey}  \& {Carswell}}{{Cooke}
  et~al.}{1997}]{cooke1997}
{Cooke} A.~J.,  {Espey} B.,   {Carswell} R.~F.,  1997, \mn@doi [\mnras]
  {10.1093/mnras/284.3.552}, \href
  {http://adsabs.harvard.edu/abs/1997MNRAS.284..552C} {284, 552}

\bibitem[\protect\citeauthoryear{{Croft}, {Weinberg}, {Katz}  \&
  {Hernquist}}{{Croft} et~al.}{1997}]{croft1997}
{Croft} R.~A.~C.,  {Weinberg} D.~H.,  {Katz} N.,   {Hernquist} L.,  1997,
  \mn@doi [\apj] {10.1086/304723}, \href
  {http://adsabs.harvard.edu/abs/1997ApJ...488..532C} {488, 532}

\bibitem[\protect\citeauthoryear{{Croft}, {Weinberg}, {Katz}  \&
  {Hernquist}}{{Croft} et~al.}{1998}]{croft1998}
{Croft} R.~A.~C.,  {Weinberg} D.~H.,  {Katz} N.,   {Hernquist} L.,  1998,
  \mn@doi [\apj] {10.1086/305289}, \href
  {http://adsabs.harvard.edu/abs/1998ApJ...495...44C} {495, 44}

\bibitem[\protect\citeauthoryear{{Dav{\'e}} \& {Tripp}}{{Dav{\'e}} \&
  {Tripp}}{2001}]{dave2001b}
{Dav{\'e}} R.,  {Tripp} T.~M.,  2001, \mn@doi [\apj] {10.1086/320977}, \href
  {http://adsabs.harvard.edu/abs/2001ApJ...553..528D} {553, 528}

\bibitem[\protect\citeauthoryear{{Dav{\'e}}, {Hernquist}, {Katz}  \&
  {Weinberg}}{{Dav{\'e}} et~al.}{1999}]{dave1999}
{Dav{\'e}} R.,  {Hernquist} L.,  {Katz} N.,   {Weinberg} D.~H.,  1999, \mn@doi
  [\apj] {10.1086/306722}, \href
  {http://adsabs.harvard.edu/abs/1999ApJ...511..521D} {511, 521}

\bibitem[\protect\citeauthoryear{{Dav{\'e}}, {Oppenheimer}, {Katz}, {Kollmeier}
   \& {Weinberg}}{{Dav{\'e}} et~al.}{2010}]{dave2010}
{Dav{\'e}} R.,  {Oppenheimer} B.~D.,  {Katz} N.,  {Kollmeier} J.~A.,
  {Weinberg} D.~H.,  2010, \mn@doi [\mnras] {10.1111/j.1365-2966.2010.17279.x},
  \href {http://adsabs.harvard.edu/abs/2010MNRAS.408.2051D} {408, 2051}

\bibitem[\protect\citeauthoryear{{Desjacques}, {Nusser}  \&
  {Sheth}}{{Desjacques} et~al.}{2007}]{vincent2007}
{Desjacques} V.,  {Nusser} A.,   {Sheth} R.~K.,  2007, \mn@doi [\mnras]
  {10.1111/j.1365-2966.2006.11134.x}, \href
  {http://adsabs.harvard.edu/abs/2007MNRAS.374..206D} {374, 206}

\bibitem[\protect\citeauthoryear{{Doroshkevich} \& {Shandarin}}{{Doroshkevich}
  \& {Shandarin}}{1977}]{doroshkevich1977}
{Doroshkevich} A.~G.,  {Shandarin} S.~F.,  1977, \mn@doi [\mnras]
  {10.1093/mnras/179.1.95P}, \href
  {http://adsabs.harvard.edu/abs/1977MNRAS.179P..95D} {179, 95P}

\bibitem[\protect\citeauthoryear{{Faucher-Gigu{\`e}re}, {Lidz}, {Hernquist}  \&
  {Zaldarriaga}}{{Faucher-Gigu{\`e}re} et~al.}{2008}]{faucher2008c}
{Faucher-Gigu{\`e}re} C.-A.,  {Lidz} A.,  {Hernquist} L.,   {Zaldarriaga} M.,
  2008, \mn@doi [\apjl] {10.1086/590409}, \href
  {http://adsabs.harvard.edu/abs/2008ApJ...682L...9F} {682, L9}

\bibitem[\protect\citeauthoryear{{Faucher-Gigu{\`e}re}, {Lidz}, {Zaldarriaga}
  \& {Hernquist}}{{Faucher-Gigu{\`e}re} et~al.}{2009}]{faucher2009}
{Faucher-Gigu{\`e}re} C.-A.,  {Lidz} A.,  {Zaldarriaga} M.,   {Hernquist} L.,
  2009, \mn@doi [\apj] {10.1088/0004-637X/703/2/1416}, \href
  {http://adsabs.harvard.edu/abs/2009ApJ...703.1416F} {703, 1416}

\bibitem[\protect\citeauthoryear{{Fern{\'a}ndez-Soto}, {Lanzetta}, {Barcons},
  {Carswell}, {Webb}  \& {Yahil}}{{Fern{\'a}ndez-Soto}
  et~al.}{1996}]{fernandez1996}
{Fern{\'a}ndez-Soto} A.,  {Lanzetta} K.~M.,  {Barcons} X.,  {Carswell} R.~F.,
  {Webb} J.~K.,   {Yahil} A.,  1996, \mn@doi [\apjl] {10.1086/309983}, \href
  {http://adsabs.harvard.edu/abs/1996ApJ...460L..85F} {460, L85}

\bibitem[\protect\citeauthoryear{{Gaikwad}, {Khaire}, {Choudhury}  \&
  {Srianand}}{{Gaikwad} et~al.}{2017a}]{gaikwad2017a}
{Gaikwad} P.,  {Khaire} V.,  {Choudhury} T.~R.,   {Srianand} R.,  2017a,
  \mn@doi [\mnras] {10.1093/mnras/stw3086}, \href
  {http://adsabs.harvard.edu/abs/2017MNRAS.466..838G} {466, 838}

\bibitem[\protect\citeauthoryear{{Gaikwad}, {Srianand}, {Choudhury}  \&
  {Khaire}}{{Gaikwad} et~al.}{2017b}]{gaikwad2017b}
{Gaikwad} P.,  {Srianand} R.,  {Choudhury} T.~R.,   {Khaire} V.,  2017b,
  \mn@doi [\mnras] {10.1093/mnras/stx248}, \href
  {http://adsabs.harvard.edu/abs/2017MNRAS.467.3172G} {467, 3172}

\bibitem[\protect\citeauthoryear{{Garzilli}, {Bolton}, {Kim}, {Leach}  \&
  {Viel}}{{Garzilli} et~al.}{2012}]{garzilli2012}
{Garzilli} A.,  {Bolton} J.~S.,  {Kim} T.-S.,  {Leach} S.,   {Viel} M.,  2012,
  \mn@doi [\mnras] {10.1111/j.1365-2966.2012.21223.x}, \href
  {http://adsabs.harvard.edu/abs/2012MNRAS.424.1723G} {424, 1723}

\bibitem[\protect\citeauthoryear{{Garzilli}, {Theuns}  \& {Schaye}}{{Garzilli}
  et~al.}{2015}]{garzilli2015a}
{Garzilli} A.,  {Theuns} T.,   {Schaye} J.,  2015, \mn@doi [\mnras]
  {10.1093/mnras/stv394}, \href
  {http://adsabs.harvard.edu/abs/2015MNRAS.450.1465G} {450, 1465}

\bibitem[\protect\citeauthoryear{{Gnedin} \& {Hui}}{{Gnedin} \&
  {Hui}}{1996}]{gnedin1996}
{Gnedin} N.~Y.,  {Hui} L.,  1996, \mn@doi [\apjl] {10.1086/310366}, \href
  {http://adsabs.harvard.edu/abs/1996ApJ...472L..73G} {472, L73}

\bibitem[\protect\citeauthoryear{{Gnedin} \& {Hui}}{{Gnedin} \&
  {Hui}}{1998}]{gnedin1998}
{Gnedin} N.~Y.,  {Hui} L.,  1998, \mn@doi [\mnras]
  {10.1046/j.1365-8711.1998.01249.x}, \href
  {http://adsabs.harvard.edu/abs/1998MNRAS.296...44G} {296, 44}

\bibitem[\protect\citeauthoryear{{Gurvich}, {Burkhart}  \& {Bird}}{{Gurvich}
  et~al.}{2017}]{gurvich2017}
{Gurvich} A.,  {Burkhart} B.,   {Bird} S.,  2017, \mn@doi [\apj]
  {10.3847/1538-4357/835/2/175}, \href
  {http://adsabs.harvard.edu/abs/2017ApJ...835..175G} {835, 175}

\bibitem[\protect\citeauthoryear{{Haardt} \& {Madau}}{{Haardt} \&
  {Madau}}{2012}]{haardt2012}
{Haardt} F.,  {Madau} P.,  2012, \mn@doi [\apj] {10.1088/0004-637X/746/2/125},
  \href {http://adsabs.harvard.edu/abs/2012ApJ...746..125H} {746, 125}

\bibitem[\protect\citeauthoryear{{Hernquist} \& {Katz}}{{Hernquist} \&
  {Katz}}{1989}]{hernquist1989}
{Hernquist} L.,  {Katz} N.,  1989, \mn@doi [\apjs] {10.1086/191344}, \href
  {http://adsabs.harvard.edu/abs/1989ApJS...70..419H} {70, 419}

\bibitem[\protect\citeauthoryear{{Hernquist}, {Katz}, {Weinberg}  \&
  {Miralda-Escud{\'e}}}{{Hernquist} et~al.}{1996}]{hernquist1996}
{Hernquist} L.,  {Katz} N.,  {Weinberg} D.~H.,   {Miralda-Escud{\'e}} J.,
  1996, \mn@doi [\apjl] {10.1086/309899}, \href
  {http://adsabs.harvard.edu/abs/1996ApJ...457L..51H} {457, L51}

\bibitem[\protect\citeauthoryear{{Hui} \& {Gnedin}}{{Hui} \&
  {Gnedin}}{1997}]{hui1997}
{Hui} L.,  {Gnedin} N.~Y.,  1997, \mn@doi [\mnras] {10.1093/mnras/292.1.27},
  \href {http://adsabs.harvard.edu/abs/1997MNRAS.292...27H} {292, 27}

\bibitem[\protect\citeauthoryear{{Jenkins} \& {Ostriker}}{{Jenkins} \&
  {Ostriker}}{1991}]{jenkins1991}
{Jenkins} E.~B.,  {Ostriker} J.~P.,  1991, \mn@doi [\apj] {10.1086/170252},
  \href {http://adsabs.harvard.edu/abs/1991ApJ...376...33J} {376, 33}

\bibitem[\protect\citeauthoryear{{Khaire} \& {Srianand}}{{Khaire} \&
  {Srianand}}{2015a}]{khaire2015b}
{Khaire} V.,  {Srianand} R.,  2015a, \mn@doi [\mnras] {10.1093/mnrasl/slv060},
  \href {http://adsabs.harvard.edu/abs/2015MNRAS.451L..30K} {451, L30}

\bibitem[\protect\citeauthoryear{{Khaire} \& {Srianand}}{{Khaire} \&
  {Srianand}}{2015b}]{khaire2015a}
{Khaire} V.,  {Srianand} R.,  2015b, \mn@doi [\apj]
  {10.1088/0004-637X/805/1/33}, \href
  {http://adsabs.harvard.edu/abs/2015ApJ...805...33K} {805, 33}

\bibitem[\protect\citeauthoryear{{Kim}, {Viel}, {Haehnelt}, {Carswell}  \&
  {Cristiani}}{{Kim} et~al.}{2004}]{kim2004}
{Kim} T.-S.,  {Viel} M.,  {Haehnelt} M.~G.,  {Carswell} R.~F.,   {Cristiani}
  S.,  2004, \mn@doi [\mnras] {10.1111/j.1365-2966.2004.07221.x}, \href
  {http://adsabs.harvard.edu/abs/2004MNRAS.347..355K} {347, 355}

\bibitem[\protect\citeauthoryear{{Kim}, {Bolton}, {Viel}, {Haehnelt}  \&
  {Carswell}}{{Kim} et~al.}{2007}]{kim2007}
{Kim} T.-S.,  {Bolton} J.~S.,  {Viel} M.,  {Haehnelt} M.~G.,   {Carswell}
  R.~F.,  2007, \mn@doi [\mnras] {10.1111/j.1365-2966.2007.12406.x}, \href
  {http://adsabs.harvard.edu/abs/2007MNRAS.382.1657K} {382, 1657}

\bibitem[\protect\citeauthoryear{{Kollmeier}, {Miralda-Escud{\'e}}, {Cen}  \&
  {Ostriker}}{{Kollmeier} et~al.}{2006}]{kollmier2006}
{Kollmeier} J.~A.,  {Miralda-Escud{\'e}} J.,  {Cen} R.,   {Ostriker} J.~P.,
  2006, \mn@doi [\apj] {10.1086/498104}, \href
  {http://adsabs.harvard.edu/abs/2006ApJ...638...52K} {638, 52}

\bibitem[\protect\citeauthoryear{{Kollmeier} et~al.,}{{Kollmeier}
  et~al.}{2014}]{kollmeier2014}
{Kollmeier} J.~A.,  et~al., 2014, \mn@doi [\apjl]
  {10.1088/2041-8205/789/2/L32}, \href
  {http://adsabs.harvard.edu/abs/2014ApJ...789L..32K} {789, L32}

\bibitem[\protect\citeauthoryear{{Kulkarni}, {Hennawi}, {O{\~n}orbe}, {Rorai}
  \& {Springel}}{{Kulkarni} et~al.}{2015}]{girish2015}
{Kulkarni} G.,  {Hennawi} J.~F.,  {O{\~n}orbe} J.,  {Rorai} A.,   {Springel}
  V.,  2015, \mn@doi [\apj] {10.1088/0004-637X/812/1/30}, \href
  {http://adsabs.harvard.edu/abs/2015ApJ...812...30K} {812, 30}

\bibitem[\protect\citeauthoryear{{Lidz}, {Faucher-Gigu{\`e}re}, {Dall'Aglio},
  {McQuinn}, {Fechner}, {Zaldarriaga}, {Hernquist}  \& {Dutta}}{{Lidz}
  et~al.}{2010}]{lidz2010}
{Lidz} A.,  {Faucher-Gigu{\`e}re} C.-A.,  {Dall'Aglio} A.,  {McQuinn} M.,
  {Fechner} C.,  {Zaldarriaga} M.,  {Hernquist} L.,   {Dutta} S.,  2010,
  \mn@doi [\apj] {10.1088/0004-637X/718/1/199}, \href
  {http://adsabs.harvard.edu/abs/2010ApJ...718..199L} {718, 199}

\bibitem[\protect\citeauthoryear{{L{\'o}pez} et~al.,}{{L{\'o}pez}
  et~al.}{2016}]{lopez2016}
{L{\'o}pez} S.,  et~al., 2016, \mn@doi [\aap] {10.1051/0004-6361/201628161},
  \href {http://adsabs.harvard.edu/abs/2016A%26A...594A..91L} {594, A91}

\bibitem[\protect\citeauthoryear{{Luki{\'c}}, {Stark}, {Nugent}, {White},
  {Meiksin}  \& {Almgren}}{{Luki{\'c}} et~al.}{2015}]{lukic2015}
{Luki{\'c}} Z.,  {Stark} C.~W.,  {Nugent} P.,  {White} M.,  {Meiksin} A.~A.,
  {Almgren} A.,  2015, \mn@doi [\mnras] {10.1093/mnras/stu2377}, \href
  {http://adsabs.harvard.edu/abs/2015MNRAS.446.3697L} {446, 3697}

\bibitem[\protect\citeauthoryear{{McDonald}}{{McDonald}}{2003}]{mcdonald2003}
{McDonald} P.,  2003, \mn@doi [\apj] {10.1086/345945}, \href
  {http://adsabs.harvard.edu/abs/2003ApJ...585...34M} {585, 34}

\bibitem[\protect\citeauthoryear{{McDonald}, {Miralda-Escud{\'e}}, {Rauch},
  {Sargent}, {Barlow}, {Cen}  \& {Ostriker}}{{McDonald}
  et~al.}{2000}]{mcdonald2000}
{McDonald} P.,  {Miralda-Escud{\'e}} J.,  {Rauch} M.,  {Sargent} W.~L.~W.,
  {Barlow} T.~A.,  {Cen} R.,   {Ostriker} J.~P.,  2000, \mn@doi [\apj]
  {10.1086/317079}, \href {http://adsabs.harvard.edu/abs/2000ApJ...543....1M}
  {543, 1}

\bibitem[\protect\citeauthoryear{{McDonald}, {Miralda-Escud{\'e}}, {Rauch},
  {Sargent}, {Barlow}  \& {Cen}}{{McDonald} et~al.}{2001}]{mcdonald2001}
{McDonald} P.,  {Miralda-Escud{\'e}} J.,  {Rauch} M.,  {Sargent} W.~L.~W.,
  {Barlow} T.~A.,   {Cen} R.,  2001, \mn@doi [\apj] {10.1086/323426}, \href
  {http://adsabs.harvard.edu/abs/2001ApJ...562...52M} {562, 52}

\bibitem[\protect\citeauthoryear{{McDonald} et~al.,}{{McDonald}
  et~al.}{2005}]{mcdonald2005}
{McDonald} P.,  et~al., 2005, \mn@doi [\apj] {10.1086/497563}, \href
  {http://adsabs.harvard.edu/abs/2005ApJ...635..761M} {635, 761}

\bibitem[\protect\citeauthoryear{{McDonald} et~al.,}{{McDonald}
  et~al.}{2006}]{mcdonald2006}
{McDonald} P.,  et~al., 2006, \mn@doi [\apjs] {10.1086/444361}, \href
  {http://adsabs.harvard.edu/abs/2006ApJS..163...80M} {163, 80}

\bibitem[\protect\citeauthoryear{{McGill}}{{McGill}}{1990}]{mcgill1990}
{McGill} C.,  1990, \mn@doi [\mnras] {10.1093/mnras/242.4.544}, \href
  {http://adsabs.harvard.edu/abs/1990MNRAS.242..544M} {242, 544}

\bibitem[\protect\citeauthoryear{{Meiksin} \& {White}}{{Meiksin} \&
  {White}}{2004}]{meiksin2004}
{Meiksin} A.,  {White} M.,  2004, \mn@doi [\mnras]
  {10.1111/j.1365-2966.2004.07724.x}, \href
  {http://adsabs.harvard.edu/abs/2004MNRAS.350.1107M} {350, 1107}

\bibitem[\protect\citeauthoryear{{Miralda-Escud{\'e}}, {Cen}, {Ostriker}  \&
  {Rauch}}{{Miralda-Escud{\'e}} et~al.}{1996}]{miralda1996}
{Miralda-Escud{\'e}} J.,  {Cen} R.,  {Ostriker} J.~P.,   {Rauch} M.,  1996,
  \mn@doi [\apj] {10.1086/177992}, \href
  {http://adsabs.harvard.edu/abs/1996ApJ...471..582M} {471, 582}

\bibitem[\protect\citeauthoryear{{Monaghan}}{{Monaghan}}{1992}]{monaghan1992}
{Monaghan} J.~J.,  1992, \mn@doi [\araa] {10.1146/annurev.aa.30.090192.002551},
  \href {http://adsabs.harvard.edu/abs/1992ARA%26A..30..543M} {30, 543}

\bibitem[\protect\citeauthoryear{{Muecket}, {Petitjean}, {Kates}  \&
  {Riediger}}{{Muecket} et~al.}{1996}]{mucket1996}
{Muecket} J.~P.,  {Petitjean} P.,  {Kates} R.~E.,   {Riediger} R.,  1996, \aap,
  \href {http://adsabs.harvard.edu/abs/1996A%26A...308...17M} {308, 17}

\bibitem[\protect\citeauthoryear{{Narayanan}, {Spergel}, {Dav{\'e}}  \&
  {Ma}}{{Narayanan} et~al.}{2000}]{narayanan2000}
{Narayanan} V.~K.,  {Spergel} D.~N.,  {Dav{\'e}} R.,   {Ma} C.-P.,  2000,
  \mn@doi [\apjl] {10.1086/317269}, \href
  {http://adsabs.harvard.edu/abs/2000ApJ...543L.103N} {543, L103}

\bibitem[\protect\citeauthoryear{{O'Meara} et~al.,}{{O'Meara}
  et~al.}{2015}]{omeara2015}
{O'Meara} J.~M.,  et~al., 2015, \mn@doi [\aj] {10.1088/0004-6256/150/4/111},
  \href {http://adsabs.harvard.edu/abs/2015AJ....150..111O} {150, 111}

\bibitem[\protect\citeauthoryear{{O'Meara}, {Lehner}, {Howk}, {Prochaska},
  {Fox}, {Peeples}, {Tumlinson}  \& {O'Shea}}{{O'Meara}
  et~al.}{2017}]{omeara2017}
{O'Meara} J.~M.,  {Lehner} N.,  {Howk} J.~C.,  {Prochaska} J.~X.,  {Fox} A.~J.,
   {Peeples} M.~S.,  {Tumlinson} J.,   {O'Shea} B.~W.,  2017, preprint, \href
  {http://adsabs.harvard.edu/abs/2017arXiv170707905O} {} (\mn@eprint {arXiv}
  {1707.07905})

\bibitem[\protect\citeauthoryear{{O'Shea}, {Nagamine}, {Springel}, {Hernquist}
  \& {Norman}}{{O'Shea} et~al.}{2005}]{oshea2005}
{O'Shea} B.~W.,  {Nagamine} K.,  {Springel} V.,  {Hernquist} L.,   {Norman}
  M.~L.,  2005, \mn@doi [\apjs] {10.1086/432645}, \href
  {http://adsabs.harvard.edu/abs/2005ApJS..160....1O} {160, 1}

\bibitem[\protect\citeauthoryear{{Padmanabhan}, {Choudhury}  \&
  {Srianand}}{{Padmanabhan} et~al.}{2014}]{hamsa2014}
{Padmanabhan} H.,  {Choudhury} T.~R.,   {Srianand} R.,  2014, \mn@doi [\mnras]
  {10.1093/mnras/stu1433}, \href
  {http://adsabs.harvard.edu/abs/2014MNRAS.443.3761P} {443, 3761}

\bibitem[\protect\citeauthoryear{{Padmanabhan}, {Srianand}  \&
  {Choudhury}}{{Padmanabhan} et~al.}{2015}]{hamsa2015}
{Padmanabhan} H.,  {Srianand} R.,   {Choudhury} T.~R.,  2015, \mn@doi [\mnras]
  {10.1093/mnrasl/slv041}, \href
  {http://adsabs.harvard.edu/abs/2015MNRAS.450L..29P} {450, L29}

\bibitem[\protect\citeauthoryear{{Palanque-Delabrouille}
  et~al.,}{{Palanque-Delabrouille} et~al.}{2015a}]{palanque2015a}
{Palanque-Delabrouille} N.,  et~al., 2015a, \mn@doi [\jcap]
  {10.1088/1475-7516/2015/02/045}, \href
  {http://adsabs.harvard.edu/abs/2015JCAP...02..045P} {2, 045}

\bibitem[\protect\citeauthoryear{{Palanque-Delabrouille}
  et~al.,}{{Palanque-Delabrouille} et~al.}{2015b}]{palanque2015b}
{Palanque-Delabrouille} N.,  et~al., 2015b, \mn@doi [\jcap]
  {10.1088/1475-7516/2015/11/011}, \href
  {http://adsabs.harvard.edu/abs/2015JCAP...11..011P} {11, 011}

\bibitem[\protect\citeauthoryear{{Peirani}, {Weinberg}, {Colombi}, {Blaizot},
  {Dubois}  \& {Pichon}}{{Peirani} et~al.}{2014}]{peirani2014}
{Peirani} S.,  {Weinberg} D.~H.,  {Colombi} S.,  {Blaizot} J.,  {Dubois} Y.,
  {Pichon} C.,  2014, \mn@doi [\apj] {10.1088/0004-637X/784/1/11}, \href
  {http://adsabs.harvard.edu/abs/2014ApJ...784...11P} {784, 11}

\bibitem[\protect\citeauthoryear{{Planck Collaboration} et~al.,}{{Planck
  Collaboration} et~al.}{2016}]{planck2016}
{Planck Collaboration} et~al., 2016, \mn@doi [\aap]
  {10.1051/0004-6361/201525830}, \href
  {http://adsabs.harvard.edu/abs/2016A%26A...594A..13P} {594, A13}

\bibitem[\protect\citeauthoryear{{Press}, {Teukolsky}, {Vetterling}  \&
  {Flannery}}{{Press} et~al.}{1992}]{press1992}
{Press} W.~H.,  {Teukolsky} S.~A.,  {Vetterling} W.~T.,   {Flannery} B.~P.,
  1992, {Numerical recipes in FORTRAN. The art of scientific computing}

\bibitem[\protect\citeauthoryear{{Puchwein}, {Bolton}, {Haehnelt}, {Madau},
  {Becker}  \& {Haardt}}{{Puchwein} et~al.}{2015}]{puchwein2015}
{Puchwein} E.,  {Bolton} J.~S.,  {Haehnelt} M.~G.,  {Madau} P.,  {Becker}
  G.~D.,   {Haardt} F.,  2015, \mn@doi [\mnras] {10.1093/mnras/stv773}, \href
  {http://adsabs.harvard.edu/abs/2015MNRAS.450.4081P} {450, 4081}

\bibitem[\protect\citeauthoryear{{Rauch} et~al.,}{{Rauch}
  et~al.}{1997}]{rauch1997}
{Rauch} M.,  et~al., 1997, \mn@doi [\apj] {10.1086/304765}, \href
  {http://adsabs.harvard.edu/abs/1997ApJ...489....7R} {489, 7}

\bibitem[\protect\citeauthoryear{{Regan}, {Haehnelt}  \& {Viel}}{{Regan}
  et~al.}{2007}]{regan2007}
{Regan} J.~A.,  {Haehnelt} M.~G.,   {Viel} M.,  2007, \mn@doi [\mnras]
  {10.1111/j.1365-2966.2006.11132.x}, \href
  {http://adsabs.harvard.edu/abs/2007MNRAS.374..196R} {374, 196}

\bibitem[\protect\citeauthoryear{{Rollinde}, {Theuns}, {Schaye}, {P{\^a}ris}
  \& {Petitjean}}{{Rollinde} et~al.}{2013}]{rollinde}
{Rollinde} E.,  {Theuns} T.,  {Schaye} J.,  {P{\^a}ris} I.,   {Petitjean} P.,
  2013, \mn@doi [\mnras] {10.1093/mnras/sts057}, \href
  {http://adsabs.harvard.edu/abs/2013MNRAS.428..540R} {428, 540}

\bibitem[\protect\citeauthoryear{{Schaye}}{{Schaye}}{2001}]{schaye2001}
{Schaye} J.,  2001, \mn@doi [\apj] {10.1086/322421}, \href
  {http://adsabs.harvard.edu/abs/2001ApJ...559..507S} {559, 507}

\bibitem[\protect\citeauthoryear{{Schaye}, {Theuns}, {Leonard}  \&
  {Efstathiou}}{{Schaye} et~al.}{1999}]{schaye1999}
{Schaye} J.,  {Theuns} T.,  {Leonard} A.,   {Efstathiou} G.,  1999, \mn@doi
  [\mnras] {10.1046/j.1365-8711.1999.02956.x}, \href
  {http://adsabs.harvard.edu/abs/1999MNRAS.310...57S} {310, 57}

\bibitem[\protect\citeauthoryear{{Schaye}, {Theuns}, {Rauch}, {Efstathiou}  \&
  {Sargent}}{{Schaye} et~al.}{2000}]{schaye2000}
{Schaye} J.,  {Theuns} T.,  {Rauch} M.,  {Efstathiou} G.,   {Sargent} W.~L.~W.,
   2000, \mn@doi [\mnras] {10.1046/j.1365-8711.2000.03815.x}, \href
  {http://adsabs.harvard.edu/abs/2000MNRAS.318..817S} {318, 817}

\bibitem[\protect\citeauthoryear{{Schaye} et~al.,}{{Schaye}
  et~al.}{2010}]{schaye2010}
{Schaye} J.,  et~al., 2010, \mn@doi [\mnras]
  {10.1111/j.1365-2966.2009.16029.x}, \href
  {http://adsabs.harvard.edu/abs/2010MNRAS.402.1536S} {402, 1536}

\bibitem[\protect\citeauthoryear{{Scoccimarro}, {Hui}, {Manera}  \&
  {Chan}}{{Scoccimarro} et~al.}{2012}]{2lpt2012}
{Scoccimarro} R.,  {Hui} L.,  {Manera} M.,   {Chan} K.~C.,  2012, \mn@doi
  [\prd] {10.1103/PhysRevD.85.083002}, \href
  {http://adsabs.harvard.edu/abs/2012PhRvD..85h3002S} {85, 083002}

\bibitem[\protect\citeauthoryear{{Shull}, {Smith}  \& {Danforth}}{{Shull}
  et~al.}{2012}]{shull2012b}
{Shull} J.~M.,  {Smith} B.~D.,   {Danforth} C.~W.,  2012, \mn@doi [\apj]
  {10.1088/0004-637X/759/1/23}, \href
  {http://adsabs.harvard.edu/abs/2012ApJ...759...23S} {759, 23}

\bibitem[\protect\citeauthoryear{{Shull}, {Moloney}, {Danforth}  \&
  {Tilton}}{{Shull} et~al.}{2015}]{shull2015}
{Shull} J.~M.,  {Moloney} J.,  {Danforth} C.~W.,   {Tilton} E.~M.,  2015,
  \mn@doi [\apj] {10.1088/0004-637X/811/1/3}, \href
  {http://adsabs.harvard.edu/abs/2015ApJ...811....3S} {811, 3}

\bibitem[\protect\citeauthoryear{{Smith}, {Hallman}, {Shull}  \&
  {O'Shea}}{{Smith} et~al.}{2011}]{smith2011}
{Smith} B.~D.,  {Hallman} E.~J.,  {Shull} J.~M.,   {O'Shea} B.~W.,  2011,
  \mn@doi [\apj] {10.1088/0004-637X/731/1/6}, \href
  {http://adsabs.harvard.edu/abs/2011ApJ...731....6S} {731, 6}

\bibitem[\protect\citeauthoryear{{Sorini}, {O{\~n}orbe}, {Luki{\'c}}  \&
  {Hennawi}}{{Sorini} et~al.}{2016}]{sorini2016}
{Sorini} D.,  {O{\~n}orbe} J.,  {Luki{\'c}} Z.,   {Hennawi} J.~F.,  2016,
  \mn@doi [\apj] {10.3847/0004-637X/827/2/97}, \href
  {http://adsabs.harvard.edu/abs/2016ApJ...827...97S} {827, 97}

\bibitem[\protect\citeauthoryear{{Springel}}{{Springel}}{2005}]{springel2005}
{Springel} V.,  2005, \mn@doi [\mnras] {10.1111/j.1365-2966.2005.09655.x},
  \href {http://adsabs.harvard.edu/abs/2005MNRAS.364.1105S} {364, 1105}

\bibitem[\protect\citeauthoryear{{Springel}, {Yoshida}  \& {White}}{{Springel}
  et~al.}{2001}]{springel2001}
{Springel} V.,  {Yoshida} N.,   {White} S.~D.~M.,  2001, \mn@doi [\na]
  {10.1016/S1384-1076(01)00042-2}, \href
  {http://adsabs.harvard.edu/abs/2001NewA....6...79S} {6, 79}

\bibitem[\protect\citeauthoryear{{Theuns} \& {Zaroubi}}{{Theuns} \&
  {Zaroubi}}{2000}]{theuns2000b}
{Theuns} T.,  {Zaroubi} S.,  2000, \mn@doi [\mnras]
  {10.1046/j.1365-8711.2000.03729.x}, \href
  {http://adsabs.harvard.edu/abs/2000MNRAS.317..989T} {317, 989}

\bibitem[\protect\citeauthoryear{{Theuns}, {Leonard}  \& {Efstathiou}}{{Theuns}
  et~al.}{1998}]{theuns1998a}
{Theuns} T.,  {Leonard} A.,   {Efstathiou} G.,  1998, \mn@doi [\mnras]
  {10.1046/j.1365-8711.1998.01740.x}, \href
  {http://adsabs.harvard.edu/abs/1998MNRAS.297L..49T} {297, L49}

\bibitem[\protect\citeauthoryear{{Theuns}, {Zaroubi}, {Kim}, {Tzanavaris}  \&
  {Carswell}}{{Theuns} et~al.}{2002}]{theuns2002}
{Theuns} T.,  {Zaroubi} S.,  {Kim} T.-S.,  {Tzanavaris} P.,   {Carswell} R.~F.,
   2002, \mn@doi [\mnras] {10.1046/j.1365-8711.2002.05316.x}, \href
  {http://adsabs.harvard.edu/abs/2002MNRAS.332..367T} {332, 367}

\bibitem[\protect\citeauthoryear{{Upton Sanderbeck}, {D'Aloisio}  \&
  {McQuinn}}{{Upton Sanderbeck} et~al.}{2016}]{sanderback2016}
{Upton Sanderbeck} P.~R.,  {D'Aloisio} A.,   {McQuinn} M.~J.,  2016, \mn@doi
  [\mnras] {10.1093/mnras/stw1117}, \href
  {http://adsabs.harvard.edu/abs/2016MNRAS.460.1885U} {460, 1885}

\bibitem[\protect\citeauthoryear{{Viel} \& {Haehnelt}}{{Viel} \&
  {Haehnelt}}{2006}]{viel2006a}
{Viel} M.,  {Haehnelt} M.~G.,  2006, \mn@doi [\mnras]
  {10.1111/j.1365-2966.2005.09703.x}, \href
  {http://adsabs.harvard.edu/abs/2006MNRAS.365..231V} {365, 231}

\bibitem[\protect\citeauthoryear{{Viel}, {Matarrese}, {Mo}, {Theuns}  \&
  {Haehnelt}}{{Viel} et~al.}{2002}]{2002MNRAS.336..685V}
{Viel} M.,  {Matarrese} S.,  {Mo} H.~J.,  {Theuns} T.,   {Haehnelt} M.~G.,
  2002, \mn@doi [\mnras] {10.1046/j.1365-8711.2002.05803.x}, \href
  {http://adsabs.harvard.edu/abs/2002MNRAS.336..685V} {336, 685}

\bibitem[\protect\citeauthoryear{{Viel}, {Haehnelt}  \& {Springel}}{{Viel}
  et~al.}{2004a}]{viel2004a}
{Viel} M.,  {Haehnelt} M.~G.,   {Springel} V.,  2004a, \mn@doi [\mnras]
  {10.1111/j.1365-2966.2004.08224.x}, \href
  {http://adsabs.harvard.edu/abs/2004MNRAS.354..684V} {354, 684}

\bibitem[\protect\citeauthoryear{{Viel}, {Weller}  \& {Haehnelt}}{{Viel}
  et~al.}{2004b}]{viel2004b}
{Viel} M.,  {Weller} J.,   {Haehnelt} M.~G.,  2004b, \mn@doi [\mnras]
  {10.1111/j.1365-2966.2004.08498.x}, \href
  {http://adsabs.harvard.edu/abs/2004MNRAS.355L..23V} {355, L23}

\bibitem[\protect\citeauthoryear{{Viel}, {Lesgourgues}, {Haehnelt}, {Matarrese}
   \& {Riotto}}{{Viel} et~al.}{2005}]{viel2005}
{Viel} M.,  {Lesgourgues} J.,  {Haehnelt} M.~G.,  {Matarrese} S.,   {Riotto}
  A.,  2005, \mn@doi [\prd] {10.1103/PhysRevD.71.063534}, \href
  {http://adsabs.harvard.edu/abs/2005PhRvD..71f3534V} {71, 063534}

\bibitem[\protect\citeauthoryear{{Viel}, {Bolton}  \& {Haehnelt}}{{Viel}
  et~al.}{2009}]{viel2009}
{Viel} M.,  {Bolton} J.~S.,   {Haehnelt} M.~G.,  2009, \mn@doi [\mnras]
  {10.1111/j.1745-3933.2009.00720.x}, \href
  {http://adsabs.harvard.edu/abs/2009MNRAS.399L..39V} {399, L39}

\bibitem[\protect\citeauthoryear{{Viel}, {Becker}, {Bolton}  \&
  {Haehnelt}}{{Viel} et~al.}{2013a}]{viel2013b}
{Viel} M.,  {Becker} G.~D.,  {Bolton} J.~S.,   {Haehnelt} M.~G.,  2013a,
  \mn@doi [\prd] {10.1103/PhysRevD.88.043502}, \href
  {http://adsabs.harvard.edu/abs/2013PhRvD..88d3502V} {88, 043502}

\bibitem[\protect\citeauthoryear{{Viel}, {Schaye}  \& {Booth}}{{Viel}
  et~al.}{2013b}]{viel2013}
{Viel} M.,  {Schaye} J.,   {Booth} C.~M.,  2013b, \mn@doi [\mnras]
  {10.1093/mnras/sts465}, \href
  {http://adsabs.harvard.edu/abs/2013MNRAS.429.1734V} {429, 1734}

\bibitem[\protect\citeauthoryear{{Viel}, {Haehnelt}, {Bolton}, {Kim},
  {Puchwein}, {Nasir}  \& {Wakker}}{{Viel} et~al.}{2016}]{viel2016}
{Viel} M.,  {Haehnelt} M.~G.,  {Bolton} J.~S.,  {Kim} T.-S.,  {Puchwein} E.,
  {Nasir} F.,   {Wakker} B.~P.,  2016, preprint, \href
  {http://adsabs.harvard.edu/abs/2016arXiv161002046V} {} (\mn@eprint {arXiv}
  {1610.02046})

\bibitem[\protect\citeauthoryear{{Webb} \& {Carswell}}{{Webb} \&
  {Carswell}}{1991}]{webb1991}
{Webb} J.~K.,  {Carswell} R.~F.,  1991, in {Shaver} P.~A.,  {Wampler} E.~J.,
  {Wolfe} A.~M.,  eds, Quasar Absorption Lines. p.~3

\bibitem[\protect\citeauthoryear{{Wiersma}, {Schaye}  \& {Smith}}{{Wiersma}
  et~al.}{2009}]{wiersma2009}
{Wiersma} R.~P.~C.,  {Schaye} J.,   {Smith} B.~D.,  2009, \mn@doi [\mnras]
  {10.1111/j.1365-2966.2008.14191.x}, \href
  {http://adsabs.harvard.edu/abs/2009MNRAS.393...99W} {393, 99}

\bibitem[\protect\citeauthoryear{{Yeche}, {Palanque-Delabrouille}, {.~Baur}  \&
  {du Mas des BourBoux}}{{Yeche} et~al.}{2017}]{yeche2017}
{Yeche} C.,  {Palanque-Delabrouille} N.,  {.~Baur} J.,   {du Mas des BourBoux}
  H.,  2017, preprint, \href
  {http://adsabs.harvard.edu/abs/2017arXiv170203314Y} {} (\mn@eprint {arXiv}
  {1702.03314})

\bibitem[\protect\citeauthoryear{{Zaldarriaga}}{{Zaldarriaga}}{2002}]{zaldarriaga2002}
{Zaldarriaga} M.,  2002, \mn@doi [\apj] {10.1086/324212}, \href
  {http://adsabs.harvard.edu/abs/2002ApJ...564..153Z} {564, 153}

\bibitem[\protect\citeauthoryear{{Zaldarriaga}, {Hui}  \&
  {Tegmark}}{{Zaldarriaga} et~al.}{2001}]{zaldarriaga2001}
{Zaldarriaga} M.,  {Hui} L.,   {Tegmark} M.,  2001, \mn@doi [\apj]
  {10.1086/321652}, \href {http://adsabs.harvard.edu/abs/2001ApJ...557..519Z}
  {557, 519}

\bibitem[\protect\citeauthoryear{{Zhan}, {Dav{\'e}}, {Eisenstein}  \&
  {Katz}}{{Zhan} et~al.}{2005}]{zhan2005}
{Zhan} H.,  {Dav{\'e}} R.,  {Eisenstein} D.,   {Katz} N.,  2005, \mn@doi
  [\mnras] {10.1111/j.1365-2966.2005.09504.x}, \href
  {http://adsabs.harvard.edu/abs/2005MNRAS.363.1145Z} {363, 1145}

\bibitem[\protect\citeauthoryear{{Zhang}, {Anninos}  \& {Norman}}{{Zhang}
  et~al.}{1995}]{zhang1995}
{Zhang} Y.,  {Anninos} P.,   {Norman} M.~L.,  1995, \mn@doi [\apjl]
  {10.1086/309752}, \href {http://adsabs.harvard.edu/abs/1995ApJ...453L..57Z}
  {453, L57}

\makeatother
\end{thebibliography}

\appendix
\vspace{-5mm}
\section{Star formation criteria}
\label{app:star-formation}
\InputFigCombine{QUICKLYA_Setting.pdf}{170}{Left, middle and right panel shows the TDR for \gthree, \ltf and \ltf with SFR criteria used in QUICK\_LYALPHA setting of \gthree. Particles with $\Delta > 1000$ and $T < 10^5$K are treated as stars and removed from further calculation. The TDR looks remarkably similar for \gthree and \ltf with QUICK\_LYALPHA setting.}{\label{fig:quick-lyalpha}}
To speed up the calculations in \gthree, we use  QUICK\_LYALPHA flag that converts gas with $\Delta > 1000$ and $T < 10^5$ K into stars. In order to study its effect on our method, we apply the same criteria to the \htf model. The left, middle and right panels in Fig. \ref{fig:quick-lyalpha} show the TDR for \gthree, \htf model without star formation criteria and \htf model with star formation criteria similar to QUICK\_LYALPHA respectively. The TDR for \gthree and \htf model with star formation are remarkably similar even at $\Delta > 1000$. We also generate \lya forest from \htf model with star formation and calculated various \lya statistics. We find that the \lya statistics are accurate to within 1.8 percent suggesting QUICK\_LYALPHA is a good approximation. This is because the particles converted in to stars occupy small volume in the simulation box. The probability that a random sightline (along which \lya optical depth is calculated) intersecting such region is small. It should be noted that the results presented in \S\ref{sec:result} for \ltf and \htf do not employ star formation criteria. 
\vspace{-5mm}
\section{Convolution of SPH kernel with Gaussian Kernel}
\label{app:sph-convolution-expression}
In this section, we show that the convolution integral in Eq. \ref{eq:smth-kern-js} can be recast in to an analytical form that is fast and easy to implement on computers. Let $W(r,h)$ be {\sc sph} kernel and $G(r,\sigma)$ be Gaussian kernel of pressure smoothing. Let $\widetilde{W}(k,h)$ and $\widetilde{G}(k,\sigma)$ be the Fourier transforms of $W(r,h)$ and $G(r,\sigma)$ respectively. The convolution of $W(r,h)$ with $G(r,\sigma)$ is given by,
\begin{equation}
\begin{aligned}
\widetilde{W^{\prime}}(k,h,\sigma) &= \widetilde{W}(k,h) \times \widetilde{G}(k,\sigma) \\
W^{\prime}(r,h,\sigma) &= \int \frac{{\rm d}^3\bm{k}}{(2 \pi)^3} \; \widetilde{S}(\bm{k},h,\sigma) \; {\rm e}^{\;i \: \bm{k} \cdot \bm{r}} \\
W^{\prime}(r,h,\sigma) &= \int \frac{{\rm d}^3\bm{k}}{(2 \pi)^3} \; \widetilde{W}(\bm{k},h) \times \widetilde{G}(\bm{k},\sigma) \; {\rm e}^{\;i \: \bm{k}  \cdot \bm{r}} \\
\end{aligned}
\end{equation}
Using the convolution theorem,
\begin{equation}
\begin{aligned}
W^{\prime}(r,h,\sigma) &=  \int \: {\rm d}^3\bm{x_1} \: W(r_1,h) \; G(|\bm{r}-\bm{x_1}|,\sigma) \;  \\
\end{aligned}
\end{equation}
The {\sc sph} kernel is given in Eq. \ref{eq:sph-kernal}. The Gaussian kernel in Eq. \ref{eq:gauss-kern-js} can be written in following form,
\begin{equation}
\begin{aligned}
G(|\bm{r}-\bm{x_1}|,\sigma) &= \frac{1}{(2 \pi \sigma^2)^{3/2}} \;\; {\rm exp} \bigg[{-\frac{|\bm{r}-\bm{x_1}|^2}{2\sigma^2}} \bigg] \\
 &= \frac{1}{(\pi b^2)^{3/2}} \;\; {\rm exp} \bigg[{-\frac{|\bm{r}-\bm{x_1}|^2}{b^2}} \bigg] \\
 &= G_0 \;\; {\rm exp} \bigg[{-\frac{|\bm{r}-\bm{x_1}|^2}{b^2}} \bigg] \\
 &= G_0 \;\; {\rm exp} \bigg[{-\frac{(r^2 + r_1^2 - 2 \: r \: r_1 \: \mu)}{b^2}} \bigg] \\
\end{aligned}
\end{equation}
where, $b = \sqrt{2} \: \sigma$ and $\mu$ is cosine of angle between vector $\bm{r}$ and $\bm{x_1}$. The convolution integral can be recast in to the following form,
\begin{multline}\label{eq:conv-approx-analytical}
W^{\prime}(r,h,\sigma) = W^{\prime}_0 \bigg[ \; \sum \limits_{n=0}^{4} {\rm A}^{+}_n \; {\rm I}_n \bigg( \frac{-r}{b}, \frac{h/2 -r}{b}\bigg) \\ + \; {\rm A}^{-}_n \; {\rm I}_n \bigg( \frac{r}{b}, \frac{h/2 +r}{b}\bigg)  + \; {\rm C}^{+}_n \; {\rm I}_n \bigg( \frac{h/2-r}{b}, \frac{h -r}{b}\bigg) \\ + \; {\rm C}^{-}_n \; {\rm I}_n \bigg( \frac{h/2+r}{b}, \frac{h +r}{b}\bigg) \bigg]
\end{multline}
where,
\begin{equation}
\begin{aligned}
W^{\prime}_0 &= \frac{\pi b^2 \: G_0 \: W_0}{r} \\
{\rm A}^{\pm}_0 &= \pm \frac{6 \: b \: r^4}{h^3} \; - \; \frac{6 \: b \: r^3}{h^2} \; + \; b\: r \\
{\rm A}^{\pm}_1 &= \frac{24 \: b^2 \: r^3}{h^3} \; \mp \; \frac{18 \: b^2 \: r^2}{h^2} \; \pm \; b^2 \\
{\rm A}^{\pm}_2 &= \pm \frac{36 \: b^3 \: r^2}{h^3} \; - \; \frac{18 \: b^3 \: r}{h^2} \\ 
{\rm A}^{\pm}_3 &= \frac{24 \: b^4 \: r}{h^3} \; - \; \frac{6 \: b^4}{h^2} \\
{\rm A}^{\pm}_4 &= \pm \frac{6 \: b^5}{h^3} \\
\end{aligned}
\end{equation}
\begin{equation}
\begin{aligned}
{\rm C}^{\pm}_0 &=\mp \frac{2 \: b \: r^4}{h^3} \; + \; \frac{6 \: b \: r^3}{h^2} \; \mp \; \frac{6 \: b \: r^2}{h} \; + \; 2\: b\: r \\ 
{\rm C}^{\pm}_1 &= -\frac{8 \: b^2 \: r^3}{h^3} \; \pm \; \frac{18 \: b^2 \: r^2}{h^2} \; - \; \frac{12 \: b^2 \: r}{h} \; \pm \; 2\: b^2 \\ 
{\rm C}^{\pm}_2 &= \mp \frac{12 \: b^3 \: r^2}{h^3} \; + \; \frac{18 \: b^3 \: r}{h^2} \; \mp \; \frac{6 \: b^3 }{h} \\ 
{\rm C}^{\pm}_3 &= -\frac{8 \: b^4 \: r}{h^3} \; \pm \; \frac{6 \: b^4}{h^2} \\ 
{\rm C}^{\pm}_4 &= \mp \frac{2 \: b^5}{h^3}  \\ 
\end{aligned}
\end{equation}
\begin{equation}
\begin{aligned}
{\rm I}_{n}(l_1,l_2) &= \int \limits_{l_1}^{l_2} \; t^n \; {\rm e}^{-t^2} \; dt \\
{\rm I}_0(l_1,l_2) &= \frac{\sqrt{\pi}}{2} \;\; \bigg[ \; {\rm erf~}(t) 
\; \bigg]_{l_1}^{l_2} \\
{\rm I}_1(l_1,l_2) &= \bigg[ \; -\frac{{\rm e}^{-t^2}}{2} \; \bigg]_{l_1}^{l_2}   \\
{\rm I}_2(l_1,l_2) &= \frac{1}{2} \; {\rm I}_0(l_1,l_2) - \frac{1}{2} \; \bigg[ t\; {\rm e}^{-t^2} \bigg] _{l_1}^{l_2} \\
{\rm I}_3(l_1,l_2) &= \;\;\;\;  {\rm I}_1(l_1,l_2) - \frac{1}{2} \; \bigg[ t^2\; {\rm e}^{-t^2} \bigg] _{l_1}^{l_2}  \\
{\rm I}_4(l_1,l_2) &= \frac{3}{2} \; {\rm I}_2(l_1,l_2) - \frac{1}{2} \; \bigg[ t^3\; {\rm e}^{-t^2} \bigg] _{l_1}^{l_2}  \\
\end{aligned}
\end{equation}

These integrals involve error function and hence need to be evaluated numerically. To speed up the calculations we used error function approximation of the following form \citep{abramowitz1972}
\begin{equation}
\begin{aligned}
{\rm erf}(x) = 1-(a_1 \: t + a_2 \: t^2 + a_3 \: t^3 + a_4 \: t^4 +a_5 \: t^5) \: {\rm e}^{-x^2} + \epsilon(x)
\end{aligned}
\end{equation}
where,
\begin{equation}
\begin{aligned}
t = \frac{1}{1 + p \: x} \;\;\;\; {\rm and} \\
|\epsilon(x)| \leq 1.5 \times 10^{-7}
\end{aligned}
\end{equation}
The small value of $|\epsilon(x)|$ indicates that the uncertainty in error function approximation is negligible. The values of the constants are
\begin{equation*}
\begin{aligned}
p &= 0.3275911 & \;\;\;\;\;\;\; & a_1 &&= 0.254829592\\
a_2 &= -0.284496736 & \;\;\;\;\;\;\; & a_3 &&= 1.421413741\\
a_4 &= -1.453152027 & \;\;\;\;\;\;\; & a_5 &&= 1.061405429\\
\end{aligned}
\end{equation*}
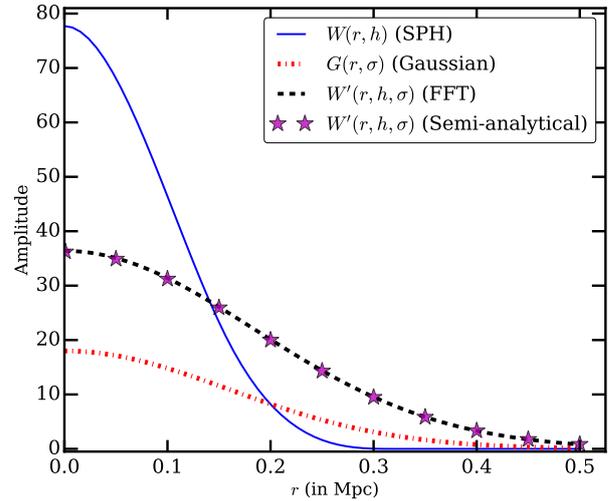
\InputFig{Analytical_Convolution.pdf}{78}{The SPH kernel (see Eq. \ref{eq:sph-kernal}) and Gaussian kernel (Eq. \ref{eq:gauss-kern-js}) for a particle are shown by blue solid curve and red dashed-dot curve respectively. These two kernels are convolve using FFT based method as shown by black dashed line. Red stars shows our the semi-analytical convolution approximation (given in Eq. \ref{eq:conv-approx-analytical}). Our method of approximation is accurate within 2 percent of FFT based method.}{\label{fig:convolution-approximation}}
Eq. \ref{eq:smth-kern-js} can also be solved numerically using a 3D FFT based method. However, we find that this method is computationally expensive for large number of particles.  Fig. \ref{fig:convolution-approximation} shows a comparison of semi-analytical approximation (red stars, given in Eq. \ref{eq:conv-approx-analytical}) with FFT based method (black dashed curve) for a particle. The SPH kernel and Gaussian kernel for this particle are shown by blue solid curve and red dash dot curves respectively. For visual purpose the Gaussian kernel is rescaled to fit the graph. Our method of approximation for convolution given in Eq. \ref{eq:conv-approx-analytical} is accurate within 2 percent of FFT based method.

\section{Jeans length in \htf model}
\label{app:jeans-length}
The \ltf and \htf models are \gtwo simulations with temperature floor $\sim 100$ K and $\sim 10000$ K respectively. For \htf model (high temperature floor) the density field would be smoother as compared to the \ltf model (low temperature floor). Hence pressure smoothing scale for \htf model would be smaller than that for \ltf. Using Eq. \ref{eq:jeans-length}, we can quantify the factor by which this pressure smoothing scale is smaller,
\begin{equation}
\begin{aligned}
L &\propto T^{\frac{1}{2} \frac{(\gamma-2)}{(\gamma-1)}} \\
\end{aligned}
\end{equation}
where we have assumed that $T \propto \Delta^{\gamma-1}$ (valid for $\Delta \leq 10$). The median \gtwo temperature (i.e., before applying {\sc cite}) in the overdensity range $\Delta \leq 10$ for \ltf and \htf model is $T_{\rm \ltf} \sim 4000$ K and $T_{\rm \htf} \sim 14000$ K respectively. The ratio of pressure smoothing scale in \ltf and \htf model is given by (assuming $\gamma \sim 1.6$),
\begin{equation}
\begin{aligned}
\frac{L_{\rm \htf}}{L_{\rm \ltf}} &= \bigg(\frac{T_{\rm \htf}}{T_{\rm \ltf}} \bigg)^{\frac{1}{2} \frac{(\gamma-2)}{(\gamma-1)}}  \approx 0.66
\end{aligned}
\end{equation}
It is clear from the above expression that if we use $L_j$ as pressure smoothing scale for the \ltf model in Eq. \ref{eq:smth-kenrel-cases} then we need to use $0.66 \times L_j$ for the \htf model. Note that we do not modify this value ($0.66$) for a different thermal history corresponding to different UVB (see Appendix \ref{sec:UVB-effect}).

\section{Effect of different path length and SNR}
\label{sec:SNR-effect}
\textbf{Effect of path length: }The analysis presented in the main paper assumes a default path length for the mock sample as $1000h^{-1}$ cMpc  (corresponding to $X \sim 5.35$ at $z=3$). With the advent of surveys like KODIAQ \citep[][$\sim 100$ QSO around $z \sim 2-3.5$]{omeara2015,omeara2017}, XQ-100 \citep{lopez2016}, the path length of \lya forest covered in QSO absorption spectroscopy is likely to increase by factor of $5$. To study the effect of increase in path length, we generated a mock sample  of 100 spectra at redshifts $z=2.5,3.0,3.5,4.0$. We repeated the procedure and generated 100 such mock samples. Thus in all we generated $5 \times 100 \times 100 = 50000$ spectra at each redshift (see section \ref{sec:method}) and followed the same procedure to calculate $8$ different statistics and the associated uncertainty (see section \ref{sec:result}). We find that the $1\sigma$ uncertainty (similar to grey shaded region in Fig.\ref{fig:delta-ps-results} to Fig. \ref{fig:b-NHI-results}) is decreased by $\sim 12$ percent but it is still dominated by sample variance. The residuals are typically less than 20 percent for \htf model. We are also able to recover the \GHI (using \htf model) within accuracy of $\sim 5$ percent although the reduced $\chi^2$ is slightly large ($\sim 1.2$) in this case  due to smaller errorbars. Thus increase in path length does not affect the results of the work presented earlier.  

\textbf{Effect of SNR:} Table \ref{tab:SNR-effect} shows the recovery of the \GTW within $1 \sigma$ statistical uncertainty from \htf model assuming \gthree as a fiducial model (fiducial \GTW$=1$) for different SNR. The path length of mock sample is $1000 h^{-1}$ cMpc (see section \ref{sec:method} for details).  The values in brackets of Table \ref{tab:SNR-effect} indicate the reduced $\chi^2$ corresponding to best fit \GTW. At all redshifts, the \htf model is able to recover the \GTW within $1\sigma$ statistical uncertainty. However, the $\chi^2_{\rm dof}$ is large for high SNR. This is because the differences between \gthree and \htf model are significant as SNR increases. But even in the case of high SNR, the reduced $\chi^2$ is close to 1 indicating the goodness of fit.
\begin{table*}
\caption{Recovery of \GTW within statistical uncertainty d \GTW for different SNR for \htf model. The path length of mock sample is $1000 h^{-1}$ cMpc ($X \sim 5.35$ at $z=3$). \gthree with $\Gamma_{12}=1$ is assumed to be fiducial model.}
\begin{tabular}{ccccc}
\hline \hline
 & \multicolumn{4}{c}{The values corresponds to \GTW $\pm$ d\GTW (best fit $\chi^2_{\rm dof}$) from \htf model} \\ \hline
SNR & $z=2.5$ & $z=3.0$ & $z=3.5$ & $z=4.0$ \\  \hline
15 & 0.97 $\pm$ 0.07 (0.96) & 1.02 $\pm$ 0.05 (0.96) & 1.00 $\pm$ 0.05 (0.70) & 0.98 $\pm$ 0.05 (0.93) \\ 
25 & 0.98 $\pm$ 0.07 (1.09) & 1.01 $\pm$ 0.05 (0.96) & 1.00 $\pm$ 0.05 (0.84) & 0.99 $\pm$ 0.05 (0.98) \\ 
35 & 0.99 $\pm$ 0.07 (1.22) & 1.02 $\pm$ 0.05 (1.07) & 0.99 $\pm$ 0.05 (0.87) & 0.97 $\pm$ 0.04 (1.03) \\ 
50 & 0.98 $\pm$ 0.07 (1.21) & 1.01 $\pm$ 0.05 (1.01) & 1.01 $\pm$ 0.05 (1.04) & 0.97 $\pm$ 0.04 (1.10) \\ 
100 & 1.02 $\pm$ 0.08 (1.33) & 1.02 $\pm$ 0.05 (1.11) & 1.00 $\pm$ 0.05 (0.95) & 0.96 $\pm$ 0.04 (1.30) \\ 
Infinite & 1.04 $\pm$ 0.07 (1.61) & 1.02 $\pm$ 0.05 (1.12) & 0.98 $\pm$ 0.05 (1.13) & 0.95 $\pm$ 0.04 (1.30) \\ \hline \hline
\end{tabular}
\label{tab:SNR-effect}
\end{table*}

\section{\lya flux statistics comparison for different thermal history}
\label{sec:UVB-effect}
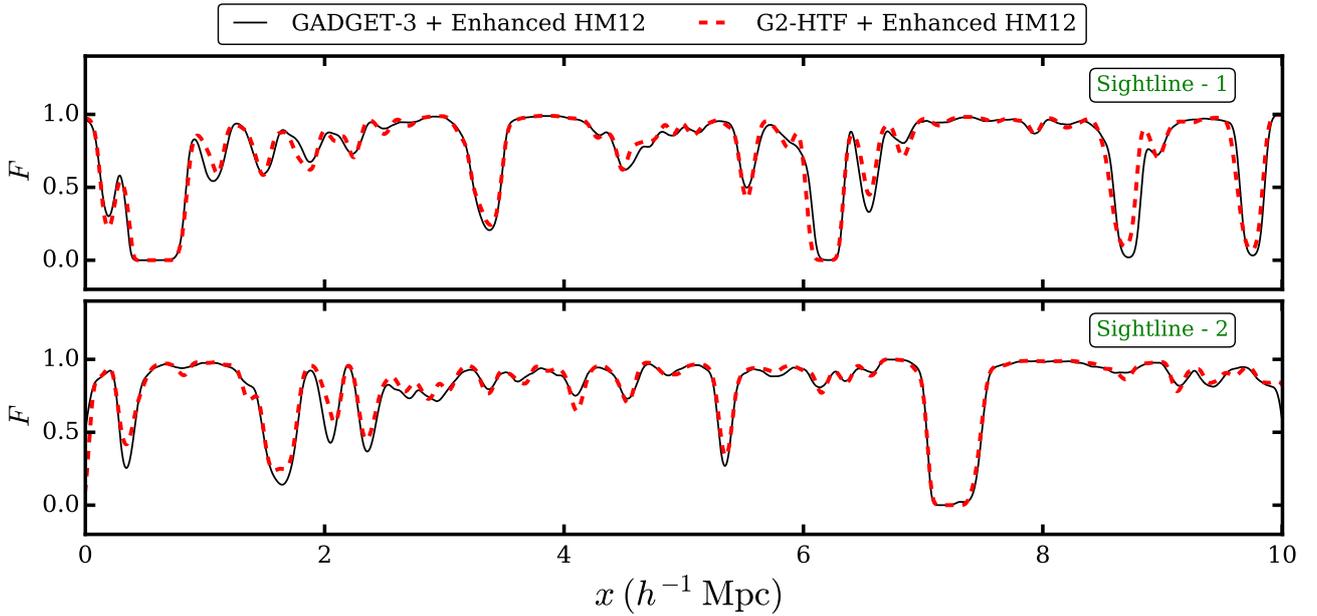
\InputFigCombine{Enhanced_HM12_LOS_Comparison.pdf}{170}{Line of sight comparison of \lya flux ($F$) for \gthree (black solid line) and \htf (red dashed line)  simulation boxes at $z=2.5$ along two different sightlines as shown in top and bottom panels. \gthree simulation is performed with an enhanced photo-heating rates (see \S \ref{sec:result} for details). For \htf model, we used enhanced HM12 photo-heating rates in {\sc cite}. The \lya flux $F$ along the sightline match very well for the two models. The \lya flux is not convolved with any LSF and no noise is added to the flux.}{\label{fig:los-comparison-enhanced-hm12}}
In order to explore the effect of difference in thermal history, we follow \citet{becker2011} and modify the photo-heating rates of species $i =$[H~{\sc i}, He~{\sc i}, He~{\sc ii}] as $\epsilon_i = a \times \epsilon_i^{\rm HM12}$ where, $\epsilon_i^{\rm HM12}$ is HM12 photo-heating rates of specie $i$. We choose $a=2.933$ such that the $T_0$ is increased by factor of $\sim 2$ while $\gamma$ remains same at all redshifts. With this updated photo-heating rates we perform a \gthree (with QUICK\_LYALPHA flag) and \htf simulation with the initial conditions same as described in \S\ref{sec:simulation}. It is important to emphasize here that we do \emph{not} perform a \gtwo simulation again, rather we only modify the HM12  photo-heating rates while running {\sc cite} in the post-processing stage on the same simulation run earlier. Note that at the initial redshift $z=6$, we use $T_0= 14543$ K and $\gamma = 1.51$ in {\sc cite} consistent with \gthree for enhanced HM12 photo-heating rates at that redshift. Fig. \ref{fig:los-comparison-enhanced-hm12} shows comparison of \lya flux from \gthree and \htf model for enhanced HM12 photo-heating rates. The flux from the two models match very well with each other. We also calculate the line of sight DPS, FPDF, FPS, wavelet PDF, curvature PDF, CDDF, $b$ parameter distribution and $b$ vs \logNHI distribution for these models. 

Figs. \ref{fig:delta-ps-results-HM12-enhanced}-\ref{fig:b-NHI-results-HM12-enhanced} show comparison of different statistics for \gthree and \htf model with enhanced HM12 UVB. 
Since the gas ionized by the enhanced HM12 UVB is at higher temperature (by a factor of $\sim 2$) as compared to the models using HM12 UVB radiation, one would expect to see the differences in the statistics. 
We notice that the \htf model residuals for FPDF are slightly large in enhanced HM12 UVB ($\sim 20$ percent, see Fig. \ref{fig:flux-pdf-results-HM12-enhanced})  as compared to those from HM12 UVB ($\sim 15$ percent, see Fig. \ref{fig:flux-pdf-results}). Similar to HM12 \htf model FPS ($\mathcal{R} \sim 5$ percent, see Fig. \ref{fig:flux-pdf-results}), we also see the mismatch in enhanced HM12 \htf model ($\mathcal{R} \sim 5$ percent, Fig. \ref{fig:flux-ps-results-HM12-enhanced}) at scales in the range $220-650$ ckpc ($k \sim 30-10 \:h$ \mpc$^{-1}$). The \htf FPS in enhanced HM12 UVB is consistent within $1.6 \sigma$ of the sample variance. Wavelets and curvature measurements are anti-correlated with temperature of the IGM. 
As expected, the comparisons of Fig. \ref{fig:wavelet-results} with Fig. \ref{fig:flux-wt-results-HM12-enhanced} and Fig. \ref{fig:flux-cs-results-HM12-enhanced} with \ref{fig:curvature-results} show that the wavelet PDF and curvature PDF are consistently smaller for enhanced HM12 UVB (higher temperature) respectively. On the other hand the $b$ parameters in enhanced HM12 UVB model (Fig. \ref{fig:bpd-results-HM12-enhanced}) are consistently larger as compared to that in HM12 UVB model (see Fig.\ref{fig:bpd-results}) at a given redshift. This trend is furthermore clear from Table \ref{tab:median-wavelet-curvature-b-enhanced-HM12} (compare with Table \ref{tab:median-wavelet}, \ref{tab:median-curvature} and \ref{tab:median-b-parameter}) where we tabulated the median value of wavelet, curvature and $b$ parameter from \gthree and \htf model with enhanced HM12 UVB. 
The wavelet and curvature PDF in enhanced HM12 UVB for \htf models are in agreement ($1.1 \sigma$) with the sample variance. Similarly the CDDF, $b$ parameter distribution and $b$ vs \logNHI lower envelope from \htf model are within sample variance with that from \gthree model (see Fig. \ref{fig:cdd-results-HM12-enhanced}, \ref{fig:bpd-results-HM12-enhanced} and \ref{fig:b-NHI-results-HM12-enhanced}).

We also calculated the reduced $\chi^2$ for different statistics. The reduced $\chi^2$ for these statistics are $\sim 0.33, 0.67, 0.34, 0.41, 0.59, 0.53, 0.47$ and $0.58$ respectively. Using this model we are also able to recover \GTW within $1\sigma$ from FPDF and FPS statistics.  This shows that \htf model is consistent (within $20$ percent) with \gthree model for a significantly different thermal history. \\ \\

\InputFigCombine{Delta_PS_Enhanced_HM12.pdf}{180}{Each panel is same as Fig. \ref{fig:delta-ps-results} except the comparison is shown for \gthree and \htf models with enhanced HM12 UVB (see the text for details).}{\label{fig:delta-ps-results-HM12-enhanced}}

\InputFigCombine{Flux_PDF_Enhanced_HM12.pdf}{180}{Each panel is same as Fig. \ref{fig:flux-pdf-results} except the comparison is shown for \gthree and \htf models with enhanced HM12 UVB (see the text for details).}{\label{fig:flux-pdf-results-HM12-enhanced}}

\InputFigCombine{Flux_PS_Enhanced_HM12.pdf}{180}{Each panel is same as Fig. \ref{fig:flux-ps-results} except the comparison is shown for \gthree and \htf models with enhanced HM12 UVB (see the text for details).}{\label{fig:flux-ps-results-HM12-enhanced}}

\InputFigCombine{Flux_WT_Enhanced_HM12.pdf}{180}{Each panel is same as Fig. \ref{fig:wavelet-results} except the comparison is shown for \gthree and \htf models with enhanced HM12 UVB (see the text for details). Median wavelet values for \gthree and \htf model are tabulated in Table. \ref{tab:median-wavelet-curvature-b-enhanced-HM12}.}{\label{fig:flux-wt-results-HM12-enhanced}}

\InputFigCombine{Flux_CS_Enhanced_HM12.pdf}{180}{Each panel is same as Fig. \ref{fig:curvature-results} except the comparison is shown for \gthree and \htf models with enhanced HM12 UVB (see the text for details). Median curvature values for \gthree and \htf model are tabulated in Table. \ref{tab:median-wavelet-curvature-b-enhanced-HM12}.}{\label{fig:flux-cs-results-HM12-enhanced}}

\InputFigCombine{CDD_Enhanced_HM12.pdf}{180}{Each panel is same as Fig. \ref{fig:cdd-results} except the comparison is shown for \gthree and \htf models with enhanced HM12 UVB (see the text for details).}{\label{fig:cdd-results-HM12-enhanced}}

\InputFigCombine{BPD_Enhanced_HM12.pdf}{180}{Each panel is same as Fig. \ref{fig:bpd-results} except the comparison is shown for \gthree and \htf models with enhanced HM12 UVB (see the text for details). Median $b$ parameter values for \gthree and \htf model are tabulated in Table. \ref{tab:median-wavelet-curvature-b-enhanced-HM12}.}{\label{fig:bpd-results-HM12-enhanced}}

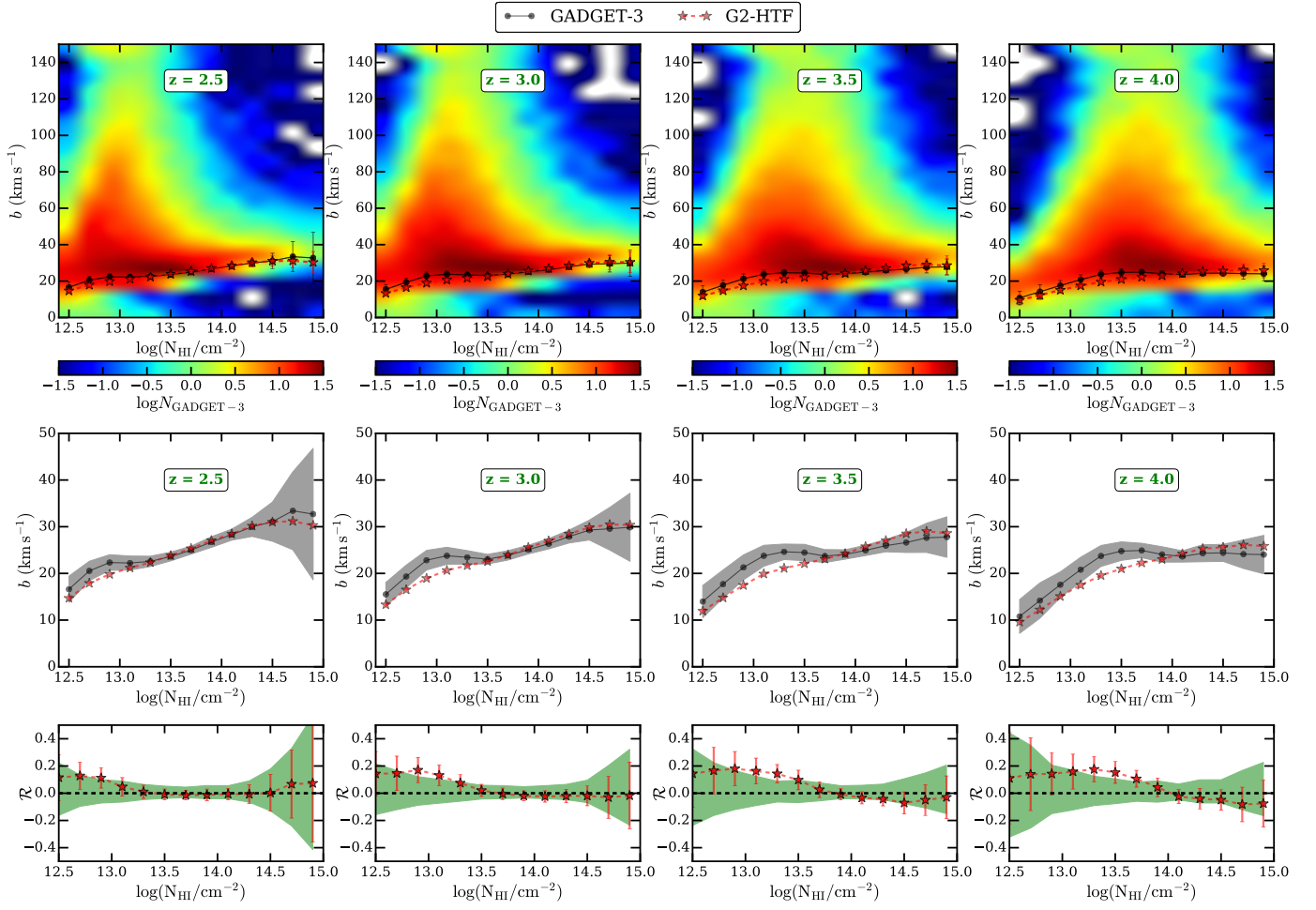
\InputFigCombine{b_vs_NHI_Enhanced_HM12.pdf}{180}{Each panel is same as Fig. \ref{fig:b-NHI-results} except the comparison in middle and bottom panel is shown for \gthree and \htf models with enhanced HM12 UVB (see the text for details).}{\label{fig:b-NHI-results-HM12-enhanced}}

\begin{table*}
\caption{Comparison of median wavelet power, curvature and $b$ parameters for the \gthree and \htf model with enhanced HM12 UVB. The errorbars correspond to 68 percentile around the median.}
\begin{center}
\begin{tabular}{ccccccc}
\hline \hline
Redshift & \multicolumn{2}{c}{Median wavelet power} & \multicolumn{2}{c}{Median curvature} & \multicolumn{2}{c}{Median $b$ parameter} \\
$z$ & \gthree & \htf & \gthree & \htf & \gthree & \htf \\  \hline 
2.5 & $-2.96 \pm 0.13$ & $-2.96 \pm 0.13$ & $-3.43 \pm 0.55$ & $-3.42 \pm 0.57$ & 32.97$^{+32.36}_{-12.20}$ & 29.50$^{+24.00}_{-11.25}$ \\ \\
3.0 & $-3.03 \pm 0.14$ & $-3.03 \pm 0.14$ & $-3.41 \pm 0.57$ & $-3.40 \pm 0.58$ & 34.22$^{+33.99}_{-12.19}$ & 30.44$^{+26.23}_{-11.15}$ \\  \\
3.5 & $-3.12 \pm 0.15$ & $-3.12 \pm 0.16$ & $-3.39 \pm 0.60$ & $-3.37 \pm 0.61$ & 36.15$^{+36.64}_{-12.67}$ & 32.23$^{+28.65}_{-11.42}$ \\ \\
4.0 & $-3.21 \pm 0.16$ & $-3.22 \pm 0.16$ & $-3.39 \pm 0.64$ & $-3.35 \pm 0.66$ & 36.76$^{+37.21}_{-13.60}$ & 33.36$^{+30.92}_{-12.31}$ \\ \hline 
\end{tabular}
\end{center}
\label{tab:median-wavelet-curvature-b-enhanced-HM12}
\end{table*}




\bsp	
\label{lastpage}
\end{document}